\documentclass[a4paper,fleqn,usenatbib]{mnras}

\usepackage[T1]{fontenc}
\usepackage{ae,aecompl}
\usepackage[caption=false,font=footnotesize]{subfig}

\usepackage{graphicx}	
\usepackage{amsmath}	
\usepackage{amssymb}	

\def\lesssim{\mathrel{\hbox{\rlap{\hbox{\lower5pt\hbox{$\sim$}}}\hbox{$<$}}}}
\def\gtrsim{\mathrel{\hbox{\rlap{\hbox{\lower5pt\hbox{$\sim$}}}\hbox{$>$}}}}
\def\mnras{MNRAS}
\def\apj{ApJ}
\def\apjl{ApJL}
\def\aj{AJ}

\def\prd{Phys.~Rev.~D}
\def\apjs{ApJS}
\def\aap{A\&A}

\def\lcdm{$\Lambda$CDM }
\def\hMpc{h^{-1} \text{Mpc}}
\def\rgm{r_{\rm gm}}
\def\hMsun{h^{-1}M_\odot}

\def\xigg{\xi_{\rm gg}}
\def\xigm{\xi_{\rm gm}}
\def\ximm{\xi_{\rm mm}}
\def\Omegam{\Omega_{\rm m}}
\def\kms{\text{kms}^{-1}}

\title[]{The effects of assembly bias on cosmological inference from galaxy-galaxy lensing and galaxy clustering.}
\author[McEwen et al.]{ 
{Joseph E. McEwen,$^{1,3}$\thanks{E-mail: mcewen.24@osu.edu}
David H. Weinberg$^{2,3}$}
\\
$^1$ Department of Physics, The Ohio State University, Columbus, Ohio 43210, USA\\
$^2$ Department of Astronomy, The Ohio State University, Columbus, Ohio
43210, USA\\
$^3$ Center for Cosmology and Astro-Particle Physics, The Ohio State
University, Columbus, Ohio 43210, USA\\
}

\date{Accepted XXX. Received YYY; in original form ZZZ}

\pubyear{2015}

\begin{document}
\label{firstpage}
\pagerange{\pageref{firstpage}--\pageref{lastpage}}
\maketitle

\begin{abstract}
The combination of galaxy-galaxy lensing (GGL) and galaxy clustering is a promising route to measuring the amplitude of matter clustering and testing modified gravity theories of cosmic acceleration.  Halo occupation distribution (HOD) modeling can extend the approach down to nonlinear scales, but galaxy assembly bias could introduce systematic errors by causing the HOD to vary with large scale environment at fixed halo mass. We investigate this problem using the mock galaxy catalogs created by Hearin \& Watson (2013, HW13), which exhibit significant assembly bias because galaxy luminosity is tied to halo peak circular velocity and galaxy colour is tied to halo formation time.  The preferential placement of galaxies (especially red galaxies) in older halos affects the cutoff of the mean occupation function $\langle N_\text{cen}(M_\text{min}) \rangle$ for central galaxies, with halos in overdense regions more likely to host galaxies. The effect of assembly bias on the satellite galaxy HOD is minimal.  We introduce an extended, environment dependent HOD (EDHOD) prescription to describe these results and fit galaxy correlation measurements.  Crucially, we find that the galaxy-matter cross-correlation coefficient, $\rgm(r)\equiv \xigm(r)\cdot [ \ximm(r) \xigg(r) ]^{-1/2}$, is insensitive to assembly bias on scales $r \gtrsim 1 \; \hMpc$, even though $\xigm(r)$ and $\xigg(r)$ are both affected individually.  We can therefore recover the correct $\ximm(r)$ from the HW13 galaxy-galaxy and galaxy-matter correlations using either a standard HOD or EDHOD fitting method.  For $M_r \leq -19$ or $M_r \leq -20$ samples the recovery of $\ximm(r)$ is accurate to 2\% or better. For a sample of red $M_r \leq -20$ galaxies we achieve 2\% recovery at $r \gtrsim 2\; \hMpc$ with EDHOD modeling but lower accuracy at smaller scales or with a standard HOD fit.  
\end{abstract}


\section{Introduction}
A central challenge of contemporary cosmology is to determine whether accelerating cosmic expansion is caused by an exotic ``dark energy" component acting within General Relativity (GR) or whether it instead reflects a breakdown of GR on cosmological scales.  One general route to distinguishing dark energy from modified gravity is to test whether the growth of structure (measured through redshift space distortions, gravitational lensing, or galaxy clustering) is consistent with GR predictions given constraints on the expansion history from supernovae, baryon acoustic oscillations (BAO), and other methods (see reviews by \citealt{2008ARA&A..46..385F,2013PhR...530...87W}).  Intriguingly, many (but not all) recent estimates of low redshift matter clustering are lower than predicted from cosmic microwave background (CMB) anisotropies evolved under a \lcdm  framework (see, \citealt{2014arXiv1401.0046M,2014arXiv1411.1074A,2015arXiv150201597P}, \lcdm = inflationary cold dark matter universe with a cosmological constant).  If this discrepancy is confirmed, it could be the first clear indication that \lcdm is an incomplete description of cosmology, and it would hint in the direction of modified gravity explanations  A promising route to measuring matter clustering is to combine galaxy clustering with galaxy-mass correlations inferred from galaxy-galaxy lensing (e.g. \citealt{2013MNRAS.432.1544M,2015ApJ...806....2M}).  This approach necessarily requires a model for galaxy bias, i.e. for the relation between galaxy and dark-matter distributions.  In this paper we examine how the theoretical uncertainties associated with modeling galaxy bias influence the matter clustering inferred from combinations of galaxy and galaxy matter clustering.

Galaxy-galaxy lensing (GGL) is a direct probe of the total matter content around a galaxy and provides a statistical relationship between galaxy and matter distributions.  Specifically GGL produces a tangential shear distortion of background galaxy images around foreground galaxies or clusters (see \citealt{2001PhR...340..291B} for a detailed review and $\S 2$ of \citealt{2013MNRAS.432.1544M} for details that are relevant to this paper).  With adequate photometric redshifts of background and foreground objects, the mean tangential shear can be converted to the projected excess surface mass density $\Delta \Sigma(R)$, where $R$ is the 2-dimensional radial distance transverse to the line of sight.  The excess surface mass density can be related to an integral of the galaxy-matter cross correlation function (\citealt{2004AJ....127.2544S})
\begin{equation} \label{Delta_Sigma} \begin{split} \Delta \Sigma(R)  & = \bar{\rho} \Big[ \frac{2}{R^2} \int_0^R \int_{-\infty}^\infty r' \xigm( r',z) dz dr' \\
& \;\;\;\;\; - \int_{-\infty}^\infty \xigm(r',z )dz \Big]~ , \end{split}  \end{equation}
with $\bar{\rho}=\Omegam \rho_{\text{crit},0}(1+z)^3 $. With sufficiently good measurements, $\Delta \Sigma(R)$ in Eqn. \ref{Delta_Sigma} can be inverted to yield the product of the mass density and galaxy-mass correlation function,  $\Omegam \xigm(r)$.  Here we will assume that this inversion can be carried out and also that the projected galaxy correlation function can be inverted to yield the 3-dimensional, real-space correlation function $\xigg(r)$ of the same galaxies for which $\Omegam \xigm(r)$ is measured by GGL. In practice, cosmological analyses may proceed by forward modeling to predict projected quantities rather than inversion to 3-d (e.g. \citealt{2013MNRAS.432.1544M,2015ApJ...806....2M,2015MNRAS.454.1161Z}).  For our present purpose of understanding the complications (potentially) caused by complex galaxy bias, it is most straightforward to focus on the 3-d quantities themselves.

The correlation functions $\xigg(r)$ and $\xigm(r)$ are related to the matter auto-correlation function $\ximm(r)$ by 
\begin{equation}
 \label{g_bias} \xigg(r)=b_\text{g}^2(r)\ximm(r)~, 
 \end{equation}
\begin{equation} 
 \label{gm_bias} \xigm(r)=b_\text{g}(r) \rgm(r) \ximm(r) ~,
 \end{equation} 
where
\begin{equation} 
\label{cross_corr}  \rgm= \frac{ \xigm }{ \sqrt{\ximm \xigg } } ~
 \end{equation}
is the galaxy-matter cross-correlation coefficient.  Equations \ref{g_bias}, \ref{gm_bias}, and \ref{cross_corr} are general and may be taken as definitions of the scale dependent bias factor $b_\text{g}(r)$ and cross correlation coefficient $\rgm(r)$.  We note that the quantity $\rgm(r)$ in real space is not constrained to be less than or equal to one in magnitude, unlike the shot-noise corrected counterpart in Fourier space (\citealt{2001MNRAS.321..439G}).  

Using Eqns. \ref{g_bias}, \ref{gm_bias}, \ref{cross_corr} one can combine observations of $\xigg(r)$ and $\Omegam \xigm(r)$ to determine
\begin{equation} 
\label{mass_inferred} \Omegam^2 \ximm(r) = \frac{ \left[ \Omegam \xigm(r) \right]^2}{ \xigg(r) } \cdot \frac{1}{\left[ \rgm(r) \right]^2 } ~. 
 \end{equation}
Thus, given a theoretical model for $\rgm(r)$, one can infer the product $\Omegam \ximm^{1/2}$, with an overall amplitude proportional to $\Omegam \sigma_8(z)$, where $\sigma_8(z)$ is the rms matter fluctuation amplitude in $8 \hMpc$ spheres at redshift $z$ and $h=H_0/100 \; \kms \;  \rm{Mpc}^{-1}$.  To a first approximation, it is the $z=0$ value of $\Omegam$ that is constrained, though for high-redshift lens and source samples the nature of the constrained parameters becomes more complex and depends on what auxiliary observational constraints are being imposed.

Under fairly general conditions, one expects $\rgm$ to approach unity on large scales, where $\xigg(r) \leq 1$ (see \citealt{2010PhRvD..81f3531B}).  However, because $\Delta\Sigma(R)$ is an integrated quantity, it is affected by small scale clustering even at large projected separation $R$ and is therefore potentially susceptible to uncertainties in non-linear galaxy bias.  To mitigate this problem, \citealt{2010PhRvD..81f3531B}, constructed a filtered GGL estimator that eliminates small scale contributions.  This approach was put in practice by \cite{2013MNRAS.432.1544M}, who applied the SDSS GGL at $R > 2 \text{ and } 4 \hMpc$ to derive constraints on $\sigma_8$ and $\Omegam$.  They found $ \sigma_8 (\Omegam/0.25)^{0.57}=0.80 \pm 0.05$, about $2 \sigma$ below the Planck+{\lcdm} prediction. 

While the method in \cite{2010PhRvD..81f3531B} and \cite{2013MNRAS.432.1544M} is already competitive with other probes of low-$z$ structure, one could do better by incorporating smaller scales, and thus increasing the signal-to-noise ratio of the GGL measurement.  This requires a description of the relation between galaxies and mass that extends to non-linear scales.  

Halo occupation distribution (HOD) modeling offers one approach to tie galaxies and dark-matter distributions down to non-linear scales (\citealt{1998ApJ...494....1J,2000MNRAS.318.1144P,2000MNRAS.318..203S,2000ApJ...531L..87M,2001ApJ...546...20S,2002ApJ...575..587B}).  The HOD specifies $P(N|M_h)$, the conditional probability that a halo of mass $M_h$ hosts $N$ galaxies of a specified class, as well as the spatial and velocity distribution of galaxies within host halos.  \cite{2006ApJ...652...26Y} showed that if one chooses HOD parameters to match galaxy clustering measurements, then the predicted GGL signal depends on the adopted cosmological model, increasing with $\sigma_8$ and $\Omegam$ in both the large scale linear regime and on smaller scales.  Several variants of the HOD modeling approach to GGL have been described in the literature (\citealt{2011ApJ...738...45L,2012PhRvD..86h3504Y,2013MNRAS.430..767C}).  \cite{2015ApJ...806....2M} measured GGL by SDSS-III BOSS galaxies (\citealt{2013AJ....145...10D}) using imaging from the CFHTlens survey (\citealt{2012MNRAS.427..146H}), and applying the methods of  \cite{2013MNRAS.430..725V} they obtained $\sigma_8=0.785^{+0.044}_{-0.044}$ for $\Omegam=0.310^{+0.019}_{-0.020}$ at the 68\% confidence interval. Recently,  \citealt{2015MNRAS.454.1161Z} have applied a modified HOD method to the SDSS main galaxy sample \citep{2002AJ....124.1810S}, obtaining an excellent joint fit to clustering and GGL for a cosmological model with $\sigma_8=0.77$ and $\Omegam=0.27$. 

The philosophy of deriving cosmological constraints from such modeling is to treat HOD quantities as ``nuisance parameters" that allow one to marginalize over uncertainties associated with galaxy formation physics (\citealt{2007ApJ...659....1Z}).   Standard HOD modeling assumes $P(N|M_h)$ is uncorrelated with the halo's large scale environment at fixed halo mass.  If $P(N|M_h)$ does depend on large scale environment, this will change the predicted galaxy clustering and galaxy-mass correlation for given set of HOD parameters.  The risk is then that modeling with an environment-independent HOD may leave systematic bias in the cosmological inferences and/or underestimate the derived cosmological parameter uncertainties associated with galaxy formation physics. 

The simplest formulation of excursion set theory \citep{1991ApJ...379..440B} predicts that halo environment is correlated with halo mass but uncorrelated with formation history at fixed mass (White 1999), motivating the idea of an environment independent HOD.  However, N-body simulations show that the clustering of halos of fixed mass varies systematically with formation time or concentration (\citealt{2004MNRAS.350.1385S,2005MNRAS.363L..66G,2006ApJ...652...71W,2006MNRAS.367.1039H,2007ApJ...657..664J}).  The dependence of halo clustering on formation time or concentration is strongest for old halos well below $M_*$.  For halos above $M_*$ there are indications that the situation is reversed (\citealt{2006ApJ...652...71W}).  In general the dependence of halo clustering on halo properties other than mass is termed \textit{assembly bias} (\citealt{2001PABei..19S..58G,2004MNRAS.350.1385S,2005MNRAS.363L..66G,2005ApJ...634...51A,2006MNRAS.367.1039H,2006ApJ...652...71W,2007MNRAS.375..633W,2007MNRAS.374.1303C,2007ApJ...654...53M,2007MNRAS.376..215B,2007ApJ...656..139W,2008MNRAS.387..921A,2008ApJ...687...12D,2009MNRAS.394.1825F,2010ApJ...708..469F}).  The physical origin of assembly bias remains unclear, though a number of explanation have been proposed.  Some level of assembly bias may arise from correlated effects of long wavelength modes on halo formation times, breaking the uncorrelated random walk assumption that underlies the minimal excursion set model.  Assembly bias can also arise in the non-linear regime from tidal truncation of low mass halo growth in the environment of high mass halos.  

If galaxy properties are tightly coupled to halo formation history, then a galaxy population can inherit assembly bias from its parent halos.  Such \textit{galaxy assembly bias} implies that $P(N|M_h)$ depends on halo environment (or halo clustering) at fixed mass.  Limited work has been carried out measuring the galactic assembly bias signal in hydrodynamic simulations.  In some simulations, the HOD has shows little to no dependence on halo environment (\citealt{2003ApJ...593....1B,2014PhDT.......126M}), which suggests that stochasticity in the galaxy formation physics in these simulations erases signatures of halo assembly bias.  However, recent work of \cite{2015arXiv150701948C} has shown a galactic assembly bias signal in the EAGLE simulation \citep{2015MNRAS.446..521S}, boosting the galaxy-correlation function by $\sim 25 \%$ on scales greater than $\sim 1 \hMpc$. Although the analysis themselves are different, the differing conclusions of \cite{2014PhDT.......126M} and \citep{2015arXiv150701948C} in simulations of volume suggest that the presence of galaxy assembly 
bias in hydrodynamic simulations depends on the adopted physical description of star formation and feedback.  Semi-analytic models predict a significant assembly bias effect in galaxy clustering for some galaxy populations (\citealt{2007MNRAS.374.1303C}), particularly red galaxies of low stellar mass. 

Abundance matching (AM) is an alternative route to populating dark-matter halos with galaxies (e.g. \citealt{2004ApJ...609...35K,2004MNRAS.353..189V,2004ApJ...614..533T,2009ApJ...696..620C,2010MNRAS.404.1111G,2009MNRAS.399..650S,2011MNRAS.414.1405N,2012ApJ...754...90W,2012ApJ...756....2R,2013ApJ...764L..31K,2015arXiv150701948C}).  Simple versions of abundance matching monotonically tie galaxy luminosity or stellar mass to some proxy for the halo or subhalo gravitational potential well, such as halo mass or maximum circular velocity.  For subhalos, AM recipes typically use $M_h$ or $V_\text{max}$ at time of accretion, with the expectation that tidal stripping will affect subhalo mass but not its stellar content \citep{2006ApJ...647..201C,2013ApJ...771...30R}.  With a subhalo mass at accretion recipe, AM is fairly successful at reproducing the galaxy content of halos in hydrodynamic cosmological simulations (\citealt{2009MNRAS.399..650S,2012MNRAS.423.3458S,2015arXiv150701948C}).  Abundance matching can easily be extended to incorporate scatter between between halo mass and galaxy properties and has been shown to be remarkably successful at reproducing observed evolution of galaxy clustering and other aspects of galaxy evolution.

Recently Hearin \& Watson (2013; hereafter HW13) have extended the AM idea to galaxy color.  Their \textit{age matching} technique monotonically maps a measure of halo formation time to galaxy color at fixed stellar mass.  Applied to the Bolshoi \lcdm \; N-body simulation  \citep{2011ApJ...740..102K}, this prescription produces good agreement with observed luminosity and color dependent clustering and GGL observations of SDSS galaxies, despite having essentially no free parameters \citep{2014MNRAS.444..729H}.  (However, \citealt{2015MNRAS.454.1161Z} show that age-matching at fixed stellar mass over predicts the GGL signal of the most luminous blue galaxies.)

By comparing clustering in the HW13 galaxy catalogs to ``scrambled" catalogs, that eliminate correlations with halo formation history, \cite{2014MNRAS.443.3044Z} show that the HW13 catalogs exhibit significant galaxy assembly bias.  For stellar mass threshold samples, this assembly bias arises because HW13 assign stellar mass based on $V_\text{max}$, and at fixed $M_h$ the halos that form earlier tend to have higher concentrations and higher $V_\text{max}$.  For color selected samples, the direct mapping between formation time and color imprints a stronger assembly bias signature.  

The HW13 catalogs adopt a physically plausible and empirically successful description of galaxy formation physics, so even if they are not correct in all details, we would like cosmological inference methods based on HOD models to be insensitive to galaxy assembly bias at this level.  

In this paper we examine the degree to which galaxy assembly bias can affect matter clustering inference results from GGL + galaxy clustering analysis.  We begin by examining the HOD and its environmental dependence in the HW13 catalogs, confirming findings of \cite{2014MNRAS.443.3044Z} but recasting them in a more HOD-specific form.  We then turn to the implications of GGL modeling, focusing our attention on the cross-correlation coefficient $\rgm(r)$, which is the quantity needed to recover $\Omegam \ximm$.  We show that HOD models fit to the galaxy correlation function of the HW13 catalogs yield accurate predictions (at the $2-5 \%$ level of precision allowed by the Bolshoi simulation volume) for $\rgm(r)$, even though they are incomplete descriptions of the bias in these galaxy populations.  We concentrate mainly on galaxy samples defined by luminosity thresholds.  However, we also consider a sample of red galaxies above a luminosity threshold, in part to examine a case with near-maximal assembly bias effects, and in part because red galaxy samples allow accurate photometric redshifts, which make them more attractive for observational GGL studies.  Our bottom line, illustrated in Fig.\ref{fig:mass_inferred_fit}, is that HOD modeling of galaxy clustering and GGL allows for accurate recovery of $\Omegam^2 \ximm(r)$ on scales $r \gtrsim 1 \hMpc$, even in the presence of galaxy assembly bias as predicted by HW13.  

\begin{figure*}
\centering
\includegraphics[width=17cm]{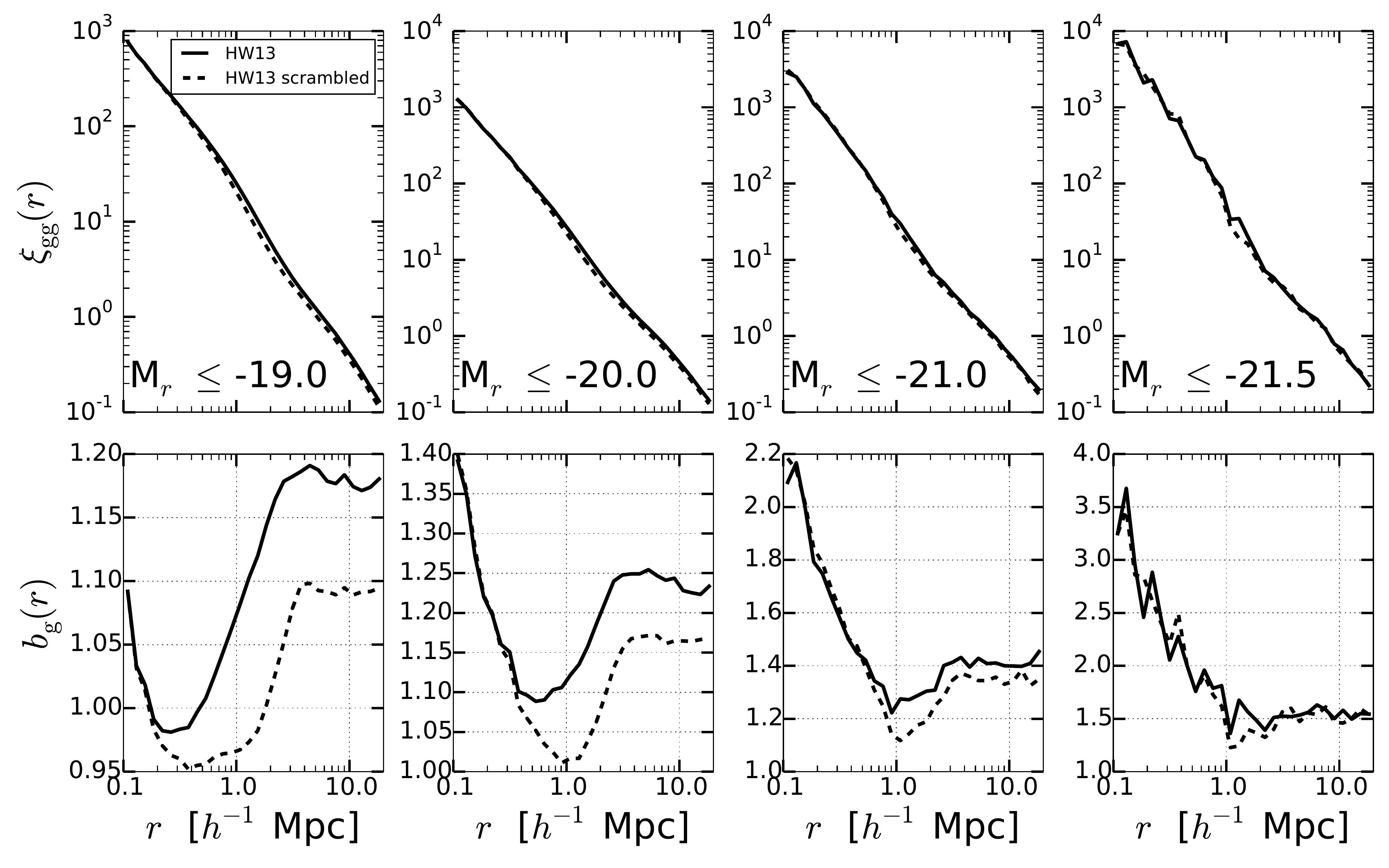}
\caption{The impact of galaxy assembly bias on the galaxy correlation function, for samples defined by four thresholds in absolute magnitude. Top panels compare $\xigg$ from the HW13 abundance matching catalog (\textit{solid}) to that of a scrambled catalog (\textit{dashed}) in which the effect of galaxy assembly bias is erased by construction.  Bottom panels plot the corresponding galaxy bias factor $b_\text{g}(r)= \sqrt{\xigg(r)/\ximm(r)}$.}
\label{fig:xi_bias}
\end{figure*}

\section{Halo Occupation Distribution of the HW13 Catalogs}

\subsection{Galaxy assembly bias in the HW13 Catalogs}
The abundance and age-matching catalogs of HW13 are built from the Bolshoi N-body simulation \citep{2011ApJ...740..102K}, which uses the Adaptive Refinement Tree (ART) code (\citealt{1997ApJS..111...73K,2008arXiv0803.4343G} ) to solve for the evolution of $2048^3$ particles in a $250 h^{-1} \text{Mpc}$ periodic box.  The mass of each particle is $m_\text{p}\approx 1.9 \times 10^8 h^{-1} M_\odot$.  The force resolution is $\epsilon \approx 1 \; h^{-1} \text{kpc}$.  The cosmological parameters are: $\Omegam$=0.27, $\Omega_\Lambda=0.73$, $\Omega_{\rm b}=0.042$, $n_s=0.95$, $\sigma_8=0.82$, and $H_0=70 \;\text{km s}^{-1} \; \text{Mpc}^{-1}$.  Bolshoi catalogs and snapshots are part of the Multidark Database and are available at {\tt{http:www.multidark.org}}.  Halos are identified by the (sub)halo finder ROCKSTAR \citep{2013ApJ...763...18B}, which uses adaptive hierarchical refinement of friends-of-friends groups in six phase-space dimensions and one time dimension.  Halos are defined within spherical regions such that the average density inside the sphere is $\Delta_\text{vir}\approx 360$ times the mean matter density of the simulation.

To create galaxy catalogs HW13 follow a two step process.  First galaxies of a particular luminosity are assigned to (sub)halos based on an abundance matching scheme.  Their abundance matching algorithm requires that the cumulative abundance of SDSS galaxies brighter than luminosity $L$ is equal to the cumulative abundance of halos and subhalos with circular velocities larger than $V_\text{max}$, $n_g(>L)=n_h(>V_\text{max})$.  Specifically HW13 uses the peak circular velocity $V_\text{peak}$ \citep{2013ApJ...771...30R}, which is the largest $V_\text{max}$ that the halo or subhalo obtains throughout its assembly history. The second step, age-matching, assigns colors by imposing a monotonic relation between galaxy colour and halo age at fixed luminosity, matching to the observed colour distribution in SDSS.  The redshift defining halo age is set to the maximum of (1) the highest redshift at which the halo mass exceeds $10^{12} \hMsun$, (2) the redshift at which the halo becomes a subhalo, (3) the redshift at which the halo's growth transitions from fast to slow accretion, as determined by the fitting function of \cite{2002ApJ...568...52W}.  Criterion (3) determines the age for most halos and subhalos.  These catalogs are publicly available at {\tt{http://logrus.uchicago.edu/{$\sim$}aphearin/}}.

When abundance matching is based on halo mass e.g. \cite{2006ApJ...647..201C}, then the resulting population of central galaxies, has no assembly bias by construction.  However, at fixed halo mass, halos that form earlier are more concentrated and thus have higher $V_\text{max}$, so even luminosity thresholded samples exhibit galaxy assembly bias in the HW13 catalogs (\citealt{2014MNRAS.443.3044Z}). We consider four samples defined by absolute magnitude thresholds $M_r-5\log h \leq -19,-20,-21,-21.5$ (hereafter we omit the $5\log h$ for brevity).  The -20 and -21 samples bracket the characteristic galaxy luminosity $L_*$, with -21 thresholds yielding the overall best clustering measurements in the SDSS main galaxy sample (\citealt{2011ApJ...736...59Z}).  The -19 threshold corresponds to fairly low luminosity galaxies with high space density, while -21.5 corresponds to rare, high luminosity galaxies.  We also consider a sample of red galaxies with $M_r \leq -20$ and $g-r \geq 0.8 -0.3(M_r + 20.0)$.   Because color is tied monotonically to halo formation time in HW13, this selection yields a near-maximal degree of galaxy assembly bias.  In addition to testing our methods under extreme conditions, this sample is observationally relevant because red galaxies allow relatively accurate photometric redshifts, making them attractive for galaxy-galaxy lensing measurements in large imaging surveys such as the Dark Energy Survey \citep{2015arXiv150705460R}. The number of galaxies in the Bolshoi simulation volume is 244766, 96595, 17250, 3954 for the $M_r \leq -19, -20, -21, $ and $-21.5$ samples, respectively, and $56591$ for the red $M_r \leq -20$ samples.  Of these, a fraction $f_\text{cen}=0.75,0.77,0.81,.85,0.77$ are central galaxies of their host halos, and a fraction $f_\text{sat}=1-f_\text{cen}$ are satellite galaxies located in subhalos.

 Fig. \ref{fig:xi_bias} compares galaxy correlation functions measured from the luminosity-threshold HW13 catalogs to those from scrambled versions of the same catalogs.  Lower panels show galaxy bias defined by $b_\text{g}(r)= \sqrt{\xigg(r)/\ximm(r)}$.  Scrambled catalogs are constructed by binning central and satellite systems in host halo mass, randomly reassigning centrals to other halos within the mass bin, then randomly reassigning satellite systems to these centrals.  By construction scrambling removes any galaxy assembly bias present in the original HW13 catalog, i.e., any correlation between the galaxy content of a halo and any halo property other than mass.  For full details of the scrambling process see \cite{2014MNRAS.443.3044Z}, who present a similar clustering analysis.  

For the $M_r \leq -19$ sample, $b_\text{g}(r)$ is about $10 \%$ higher in the HW13 catalog relative to the scrambled catalogs at $r > 3\; \hMpc$.  Both catalogs show a drop in $b_\text{g}(r)$ as $r$ decreases from $3 \hMpc$ to $0.5 \hMpc$, then a rise on still smaller scales.  The two correlation functions converge at $r < 0.5\; \hMpc$.  

The $M_r \leq -20$ sample shows similar behavior, but the differences between scrambled and unscrambled correlation functions are somewhat smaller.  For luminous, $M_r \leq -21$ galaxies, the difference in the large scale bias is only $\sim 3 \%$, and the two correlation functions are essentially converged at $r \lesssim 1 \; \hMpc$.  For the most luminous sample, $M_r \leq -21.5$, any differences are smaller still, and consistent with noise in $\xigg(r)$. 

The trends in Fig. \ref{fig:xi_bias} make sense in light of the previous studies of halo assembly bias, which show that the dependence of clustering on formation time is strongest for low mass halos and declines as the halo mass approaches the characteristic mass $M_*$ of the halo mass function \citep{2005MNRAS.363L..66G,2006MNRAS.367.1039H,2006ApJ...652...71W}. The minimum halo mass for $M_r \leq -19$ galaxies is low, and halos in denser environments are more likely to host HW13 galaxies because they have earlier formation times and higher circular velocities at fixed mass.  This preferential formation in dense environments accounts for the higher large scale bias factor of the HW13 catalog relative to the scrambled catalog.  As the luminosity threshold and minimum host halo mass increase, the bias factor grows but the impact of assembly bias diminishes.  For $M_r \leq -21.5$ the minimum halo mass is $M_h \approx 10^{13} \hMsun$ (see Fig. \ref{fig:rest_HOD} below), and any residual impact of assembly bias on the galaxy population is no longer discernible. 

In the lower luminosity samples, $b_\text{g}(r)$ becomes scale-dependent (at the $10-20 \%$ level) at the transition between the 2-halo regime of $\xigg(r)$, where galaxy pairs come from separate halos, and the 1-halo regime dominated by galaxy pairs within a single halo.  Any halo massive enough to contain two galaxies is far above the minimum mass threshold for a central galaxy, so any assembly bias effects in the 1-halo regime will arise from the satellite galaxy population.  The convergence of correlation functions at small $r$ suggests that assembly bias effects in the HW13 catalog are driven by the central galaxy population rather than satellites, a point we demonstrate explicitly in below.

\begin{figure}
\centering
\includegraphics[width=8.5cm]{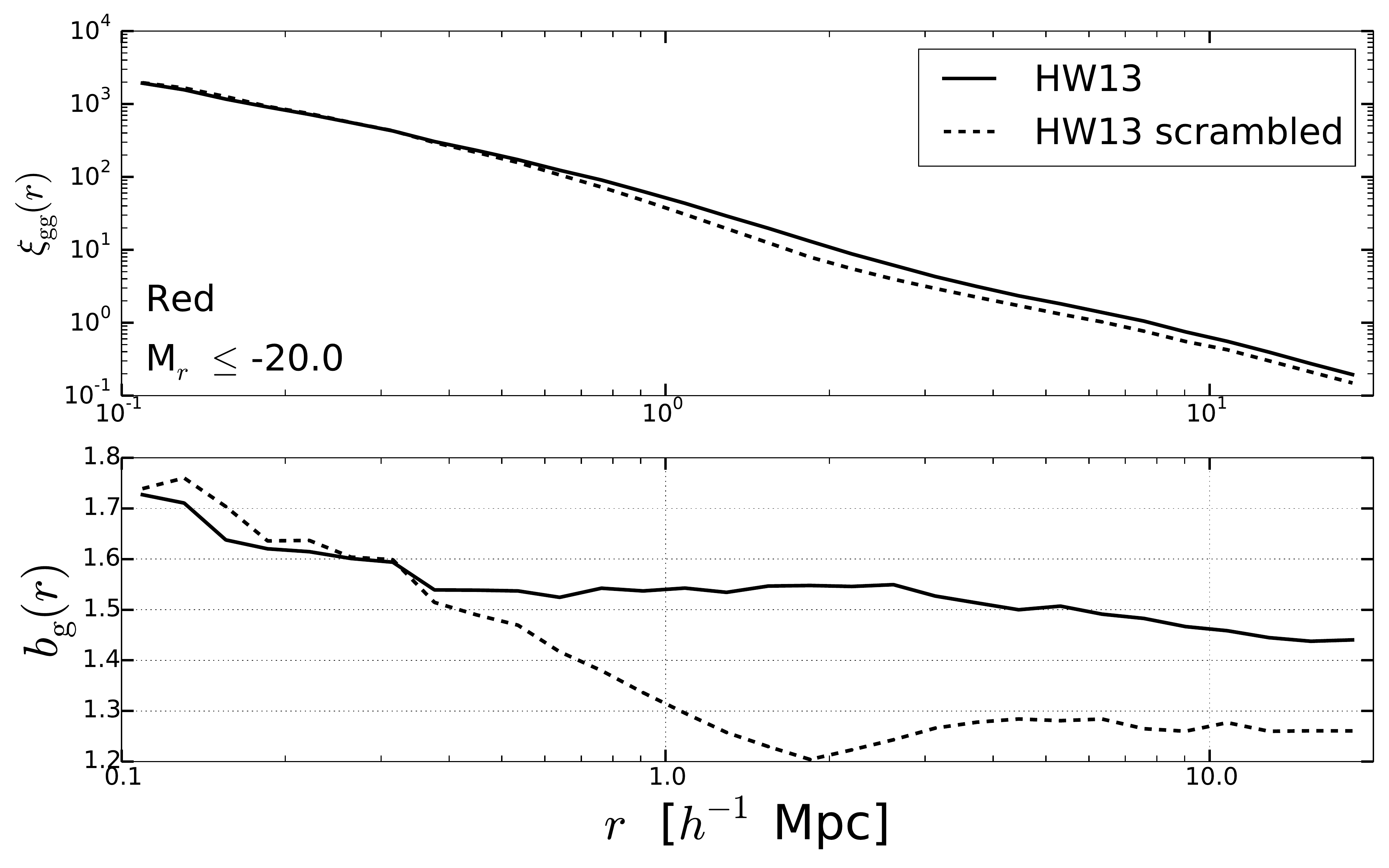}
\caption{Galaxy assembly bias in $M_r \leq -20$ red samples.  As in Fig. \ref{fig:xi_bias}, the top panel compares the measured galaxy correlation function in HW13 to a scrambled version of HW13, and the bottom panel compares results for the galaxy bias factor.  }
 \label{fig:xi_bias_red}
\end{figure}

Figure \ref{fig:xi_bias_red} shows galaxy correlation functions and galaxy bias results for red $M_r \leq -20$ galaxies (our maximal galaxy assembly bias sample), again comparing HW13 to scrambled catalogs.  The large difference in bias factors in the bottom panel of Fig. \ref{fig:xi_bias_red} is indicative of the strong galaxy assembly bias for low luminosity red galaxies in HW13.  As with the luminosity threshold case, $\xigg(r)$ for colour selected HW13 and scrambled catalogs converges on small scales, indicating the assembly bias is also primarily due to central galaxy populations. 

\begin{figure}
\centering
\includegraphics[width=8.5cm]{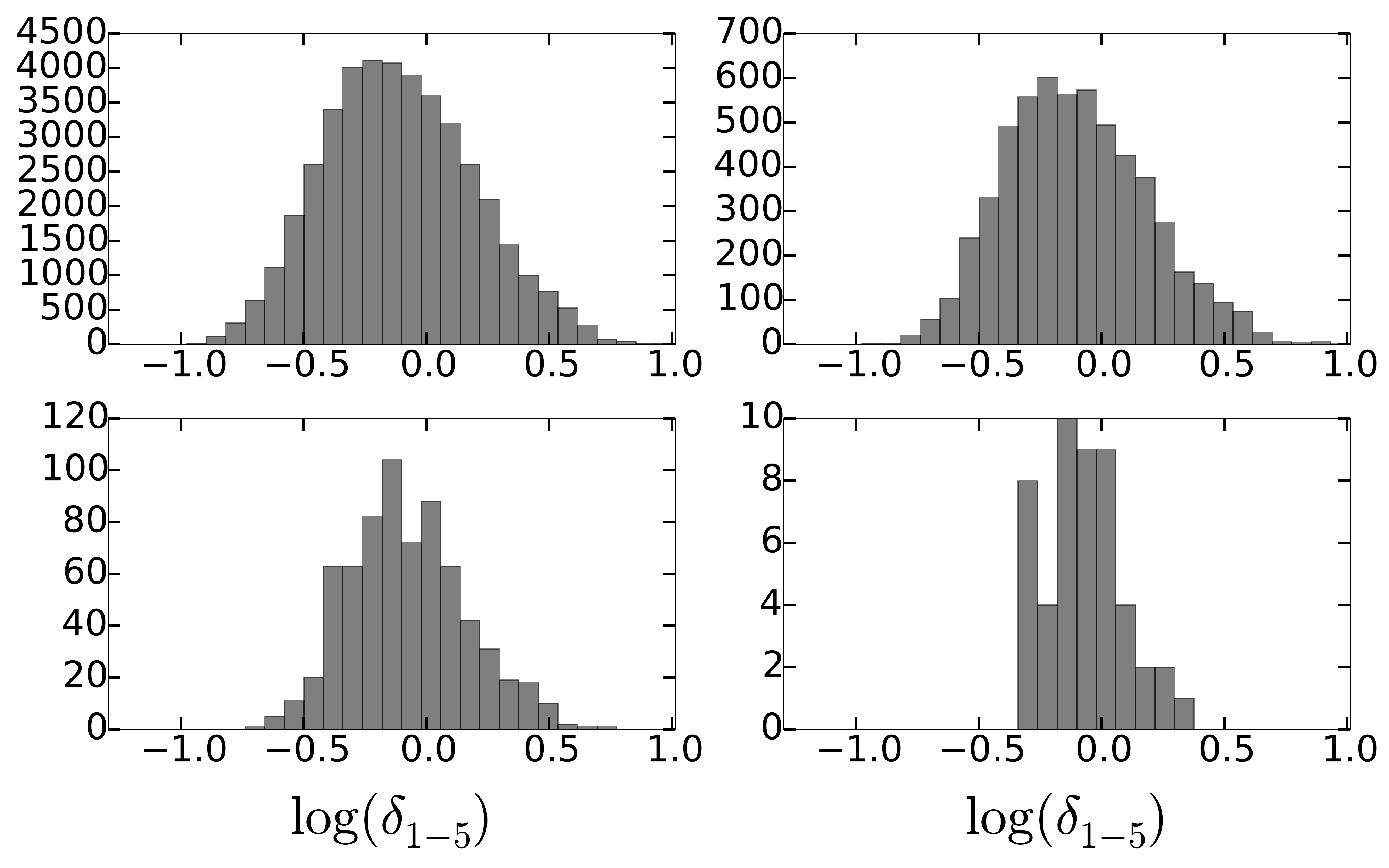}
\caption{Distribution of halo environments for halos in four $0.1$-dex bins of mass, centered at $\log M_h/ \hMsun=11.25, 12.25, 13.25,$ and $14.25$ (\textit{top left to bottom right}).  The density contrast $\delta$ is measured in a spherical annulus of $1 < r < 5 \hMpc$.} 
\label{fig:ED_dist}
\end{figure}

\subsection{HOD Analysis of $M_r \leq -19$ Galaxies}

The HOD specifies the probability $P(N | M_h)$ that a halo of mass $M_h$ contains $N$ galaxies of a specified class, together with auxiliary prescriptions that specify the spatial and velocity distributions of galaxies within halos \citep{2000MNRAS.311..793B,2002ApJ...575..587B}.  Following \cite{2001MNRAS.321..439G} and \cite{2004ApJ...609...35K}, we separate $P(N | M_h)$ into contributions from central and satellite galaxies, 
\begin{equation} P(N|M_h) = P(N_\text{cen} | M_h) + P(N_\text{sat} |M_h) ~. \end{equation} 
We adopt the parameterization of \cite{2005ApJ...633..791Z}, which provides a good fit to the theoretical predictions of hydrodynamic simulations and semi-analytic models and a good fit to observed galaxy correlation functions (\citealt{2005ApJ...630....1Z,2011ApJ...736...59Z,2012A&A...542A...5C,2015MNRAS.449.1352C}).

The mean occupation of central galaxies is described by 
\begin{equation} \label{N_cen}\begin{split}  \langle N_\text{cen} \rangle & = \frac{1}{2}\Big[ 1-  \text{erf} \left( \frac{ \log M_h - \log M_\text{min} }{ \sigma_{\log M_h}  } \right) \Big] ~.  \end{split}  \end{equation}  
The parameter  $M_\text{min}$ sets the scale where $\langle N_\text{cen} \rangle=1/2$ while $\sigma_{\log M_h}$ controls the sharpness of the transition from $\langle N_\text{cen} \rangle=0$ to $\langle N_\text{cen} \rangle=1$.  Physically $\sigma_{\log M_h}$ represents the scatter between halo mass and central galaxy luminosity.  A large scatter corresponds to a soft transition and a small scatter to a sharp transition. 

Satellite occupancy is determined by 
\begin{equation} \label{N_sat} \langle N_\text{sat} \rangle = \left( \frac{ M_h}{M_1} \right)^\alpha \exp\left( - \frac{M_\text{cut}}{M_h} \right) ~. \end{equation}
The HOD parameter $M_1$ is approximately the mass scale at which halos have an average of one satellite.  At larger halo masses the satellite occupancy increases as a power-law with slope $\alpha$.   $M_\text{cut}$ controls the scale at which the power law is truncated at low mass. 

The total mean occupancy is the sum of central and satellite mean occupancies
\begin{equation} \langle N \rangle = \langle N_\text{cen} \rangle + \langle N_\text{sat} \rangle ~. \end{equation} 
A host halo is assigned a central galaxy by Bernoulli sampling with Eqn. \ref{N_cen} serving as the probability for success.  If a host halo is determined to contain a central, the number of satellites assigned to the central is done by Poisson sampling with Eqn. \ref{N_sat} serving as the average. 

We want to examine the dependence of the HOD on the large scale environment of halos at fixed $M_h$. We define halo environment by the dark matter density contrast $\delta_{1-5}$ measured in a spherical annulus of $1 < r < 5 \hMpc$.  Figure \ref{fig:ED_dist} shows the distribution of $\delta_{1-5}$ in four narrow bins of log $M_h$.  As expected, the higher mass halos tend to reside in higher density regions, giving rise to the well known mass dependence of halo bias.  To remove the trend from our analysis, we rank the halos by $\delta_{1-5}$ in narrow (0.2-dex) mass bins, so we can compare the HOD of halos in, e.g., the 20\% highest or lowest density environment relative to other halos of nearly equal mass.  We have experimented with different definitions of environment and found that our overall results are insensitive to, e.g., changing the radii of the spherical annulus, including the central $1 \hMpc$, or incorporating distance to nearest large halo as an environmental measure. 

\begin{figure}
\centering
\includegraphics[width=8.5cm]{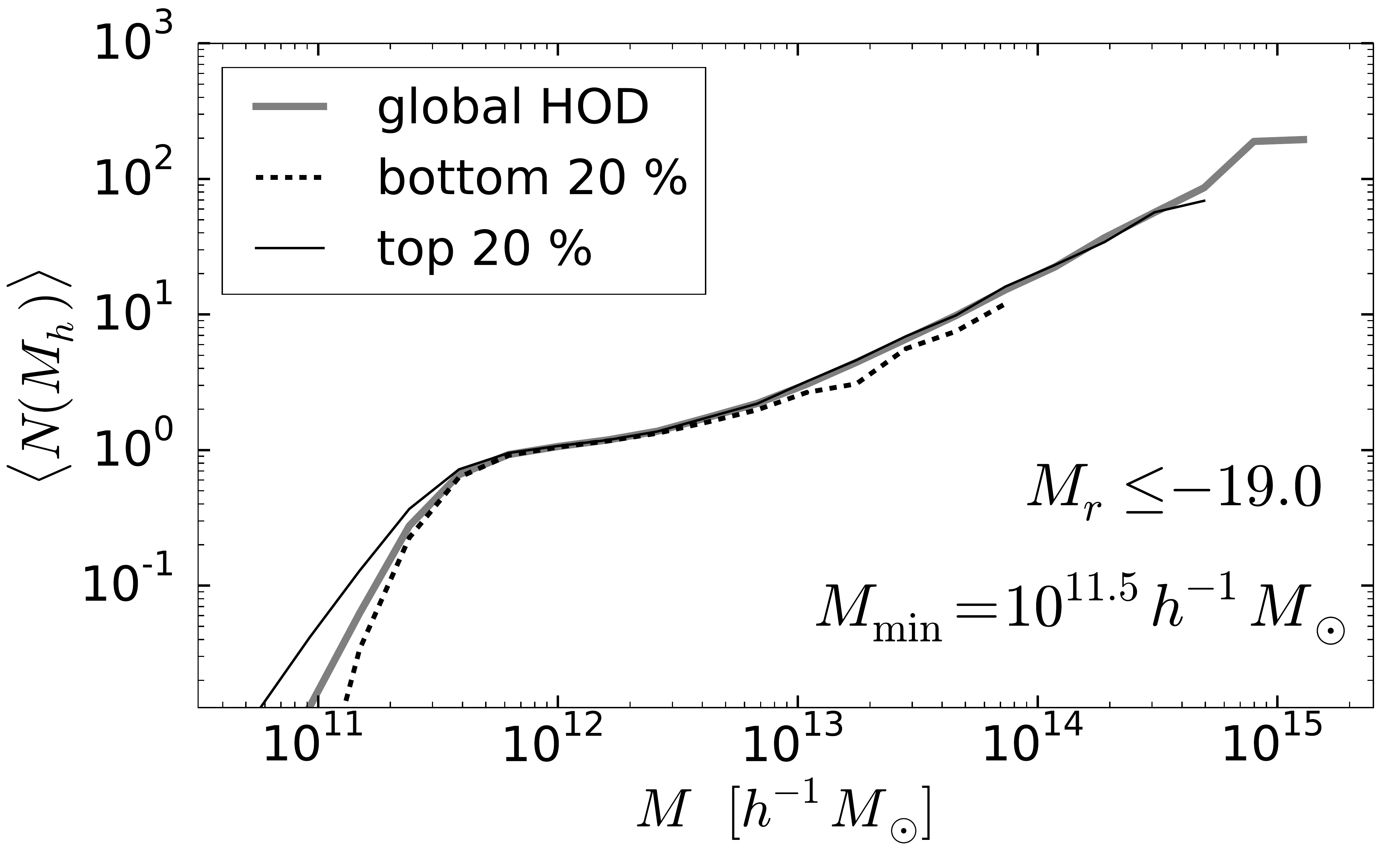}
\caption{The Measured HOD in the HW13 catalogs for the $M_r \leq -19$ sample. The grey line shows the global mean occupation function $\langle N(M_h) \rangle$ for halos in all environments.  Solid and dashed black curves show $\langle N(M_h)\rangle$ for halos in the $20 \%$ highest and lowest density environments, respectively, as measured by $\delta_{1-5}$.  For the global HOD, $\langle N(M_h) \rangle =0.5$ at $M_h=M_\text{min}=10^{11.5}\;  \hMsun$. }
\label{fig:19_HOD}
\end{figure}

Fig. \ref{fig:19_HOD} illustrates the HOD dependence on host halo environment in the HW13 catalog.  The solid grey curve shows $\langle N (M_h) \rangle$ for the global HOD computed by counting galaxies in 0.2 dex bins of $M_h$ with out reference to environment.  Solid (dashed) curves show $\langle N(M_h) \rangle$ computed for the 20 \% of halos with the highest (lowest) density environments in each mass bin.  The shape of the measured HOD curves in Fig. \ref{fig:19_HOD} is similar to the functional form predictions of Eqns. \ref{N_cen} and \ref{N_sat}: a sharp rise in $\langle N \rangle$ from zero to one associated with central galaxies, and a shallow plateau between $\langle N \rangle =1-2$, followed by a steepening to a power law.  The environmental dependence is visually evident, primarily for low mass host halos where $ \langle N \rangle < 1$.  In the language of HOD parameters, the HOD for halos in higher density environments (top 20 \%) has a lower $M_\text{min}$ and a larger $\sigma_{\log M_h}$ than the global HOD, and the reverse is true in low density environments.  Halos with $M_h \sim 1-2 \times 10^{11} \hMsun$ are therefore more likely to host a central galaxy with $M_r \leq -19$ if they reside in a high density environment, giving rise to the higher bias factor seen in Fig. \ref{fig:xi_bias}, relative to the scrambled catalog which has the environment independent, global HOD by construction.  The satellite $\langle N(M_h) \rangle$ shows little dependence on environment, though the highest mass halos are only present in the dense environments. 

How much of the assembly bias in the HW13 catalog is explained by this dependence of $\langle N (M_h) \rangle $ on the $5 \hMpc$ environment?  To answer this question, we construct catalogs with an environment-dependent HOD (EDHOD) and compare their clustering to that of the HW13 galaxies.  As a first step, we measure $\langle N(M_h) \rangle$ in 30 bins of environment $\delta_{1-5}$ (and the same 0.2-dex mass bins).  This bin-wise EDHOD automatically incorporates environmental dependence for both central and satellite galaxies. In the terminology introduced by \cite{2015arXiv151203050H} one can regard our EDHOD as a ``decorated HOD" with $\delta_{1-5}$ as the additional control variable. 

\begin{figure}
\centering
\includegraphics[width=8cm]{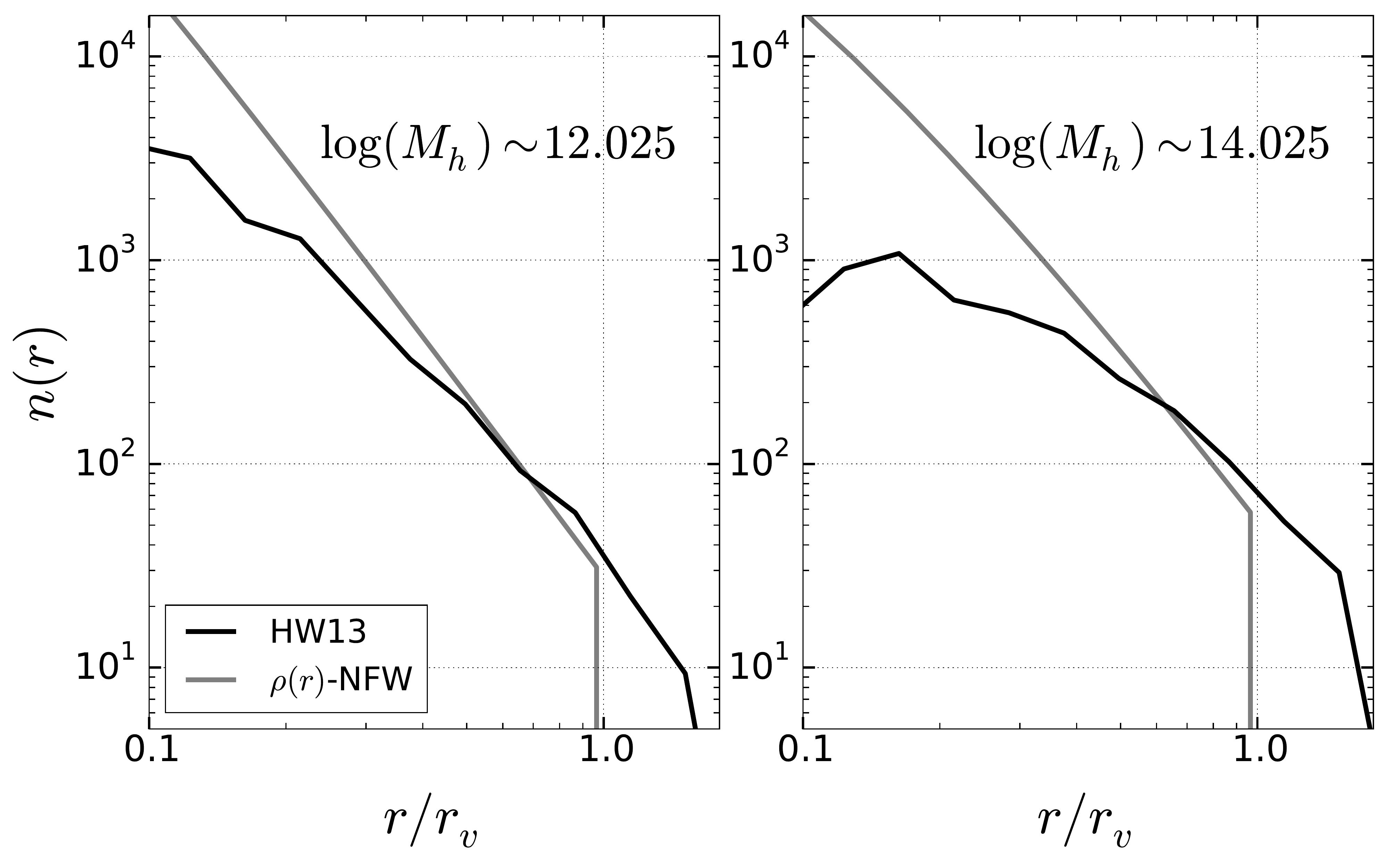}
\caption{ Radial distributions of HW13 $M_r \leq -19$ satellite galaxies(\textit{solid black curves}), in halos with $\log M_h/\hMsun= 12-12.05$ (\textit{left}) and $\log M_h/\hMsun= 14-14.05$ (\textit{right}). Grey curves show an NFW profile with the mean concentration expected for this halo mass truncated at the viral radius.}
\label{fig:sat_dist}
\end{figure}

After choosing the number of central and satellite galaxies in each halo by drawing from $P(N|M_h, \delta_{1-5})$, we must determine the positions of galaxies within the halos. A standard approach is to place the central galaxy at the halo center-of-mass and distribute satellite galaxies with a \cite{1997ApJ...490..493N} type profile (hereafter NFW) so that $ n(r) \propto \rho_\text{NFW}(r)$.  However, we find that the satellite galaxy distribution in the HW13 catalogs differs substantially from an NFW profile, a consequence of satellites being placed within subhalos in the AM scheme (\citealt{2005ApJ...618..557N,2005ApJ...624..505Z}).  Figure \ref{fig:sat_dist} illustrates this difference, comparing the HW13 satellite profiles in two narrow bins of halo mass to an NFW profile. The radial profile of HW13 satellites is much flatter than an NFW profile, and it extends beyond the viral radius because Rockstar halos are aspherical while $r_v$ is defined with a spherical overdensity.  We therefore use the measured HW13 radial profiles rather than an NFW form to create our EDHOD catalogs.  We found that clustering on scales $r \leq 2 \hMpc$ would be substantially different if we imposed an NFW profile for the satellite distribution. 

\begin{figure}
\centering
\includegraphics[width=8.5cm]{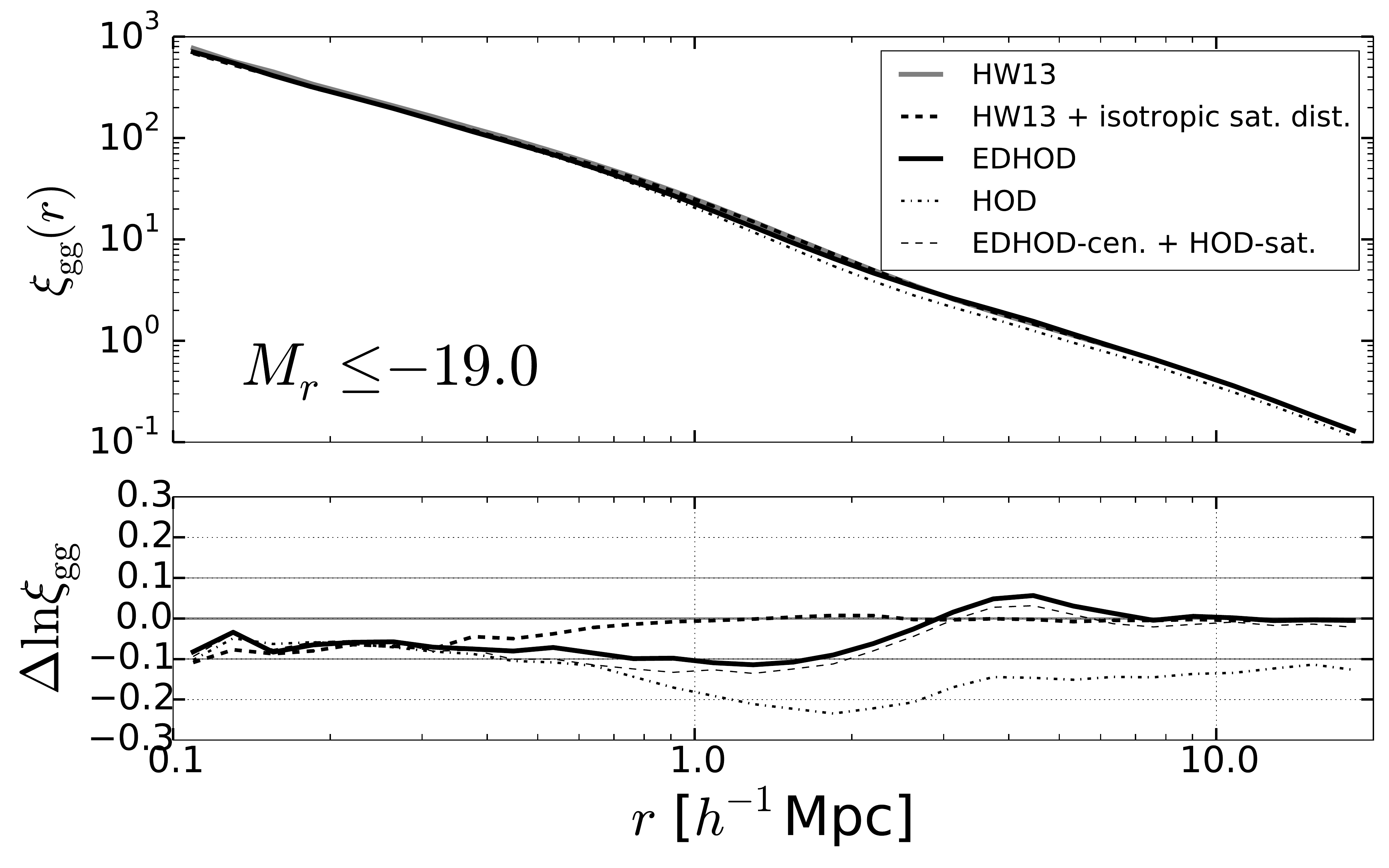}

\caption{Galaxy-correlation function for the $M_r \leq -19$ HW13 catalog compared to several HOD realizations.  The grey curve, obscured in the upper panel, shows $\xigg(r)$ from the HW13 catalog.  Dot-dashed and solid black curves show $\xigg(r)$ from catalogs created using the global HOD and environmentally dependent HOD (EDHOD), respectively, measured from the HW13 catalog.  The bottom panel shows fractional deviations from the HW13 $\xigg(r)$.  Additional curves show the effect of isotropizing the satellite distributions in the HW13 catalog (\textit{heavy dashed}) or of combining the environmentally dependent HOD for centrals with the global HOD for satellites (\textit{light dashed}).}
\label{fig:HOD_compare}
\end{figure}

In Fig. \ref{fig:HOD_compare} we compare galaxy auto-correlation results taken from our measured (ED)HOD catalogs and from HW13 for $M_r \leq -19$ samples.  The top panel plots $\xigg(r)$ while the bottom panel shows the fractional difference in $\xigg(r)$ compared to HW13.  As seen in the lower panel, randomizing the host-centric satellite angles in the HW13 catalog, thus removing the effects of halo ellipticity and substructure, depresses $\xigg(r)$ by up to 10\% below $1 \; \hMpc$ but has negligible effect at larger separations.  The standard HOD model underpredicts the HW13 $\xigg(r)$ even at large separations, an indication of the impact of galaxy assembly bias, as already seen in Fig. \ref{fig:xi_bias}.  The EDHOD model, on the other hand, matches HW13 almost perfectly at $r > 7 \; \hMpc$.   However, the EDHOD $\xigg(r)$ rises 5\% above HW13 at $r \approx 5 \; \hMpc$ and falls 10\% below at $r \approx 1-2 \; \hMpc$, before converging to the isotropized satellite case at still smaller scales.  The dashed curve shows the effect of imposing an EDHOD for central galaxies but using the global HOD for satellites. These results are nearly identical to those of the full EDHOD model, demonstrating that for this sample it is central galaxy environment dependence that matters.  We conclude that incorporating the environmental dependence of central galaxy occupations reproduces the large scale bias of the HW13 catalog but leaves a 5-10\% residual in $\xigg(r)$ on non-linear scales. 

\begin{figure}
\centering
\includegraphics[width=8cm]{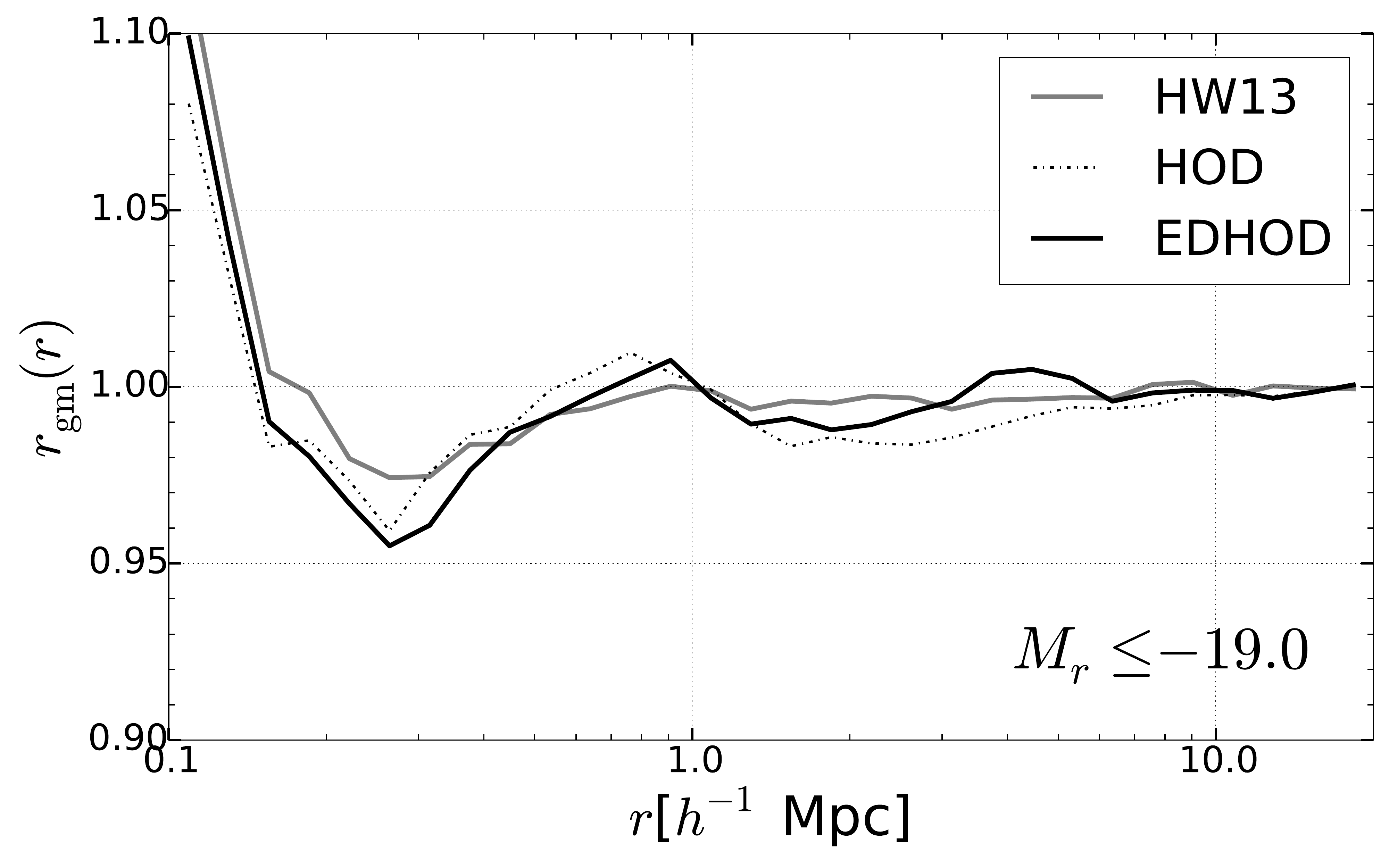}
\caption{Galaxy-matter cross-correlation coefficient (Eqn. \ref{cross_corr}) computed from the HW13 $M_r \leq -19$ catalog (\textit{grey}) or from catalogs created by using the global HOD (\textit{black solid}) or EDHOD (\textit{dot-dashed}) of this sample.}
\label{fig:19_r_true}
\end{figure}

Figure \ref{fig:19_r_true} plots the $M_r \leq -19$ cross correlation coefficient $\rgm(r)$ for the HW13, HOD, and  EDHOD catalogs.  The HW13 curve remains  close to unity (within 0.5\%) at $r > 1\; \hMpc$, with a drop and rise inside $0.4\; \hMpc$.  The EDHOD prediction is strikingly similar, matching HW13 to 1\% or better at $r > 0.4 \; \hMpc$ and showing similar form at smaller scales.  Even the global HOD prediction is similar, deviating by 1.5\% at $r > 0.4 \; \hMpc$, despite the much larger deviation in $\xigg(r)$ seen in Fig. \ref{fig:19_HOD}.  These results are our first indication that using a standard HOD versus an environment-dependent HOD has little impact on matter clustering inferences. 

\subsection{Results for other Galaxy Samples}

\begin{figure}
\captionsetup[subfigure]{labelformat=empty}
\centering
	\subfloat[]{
   \includegraphics[width=8cm]{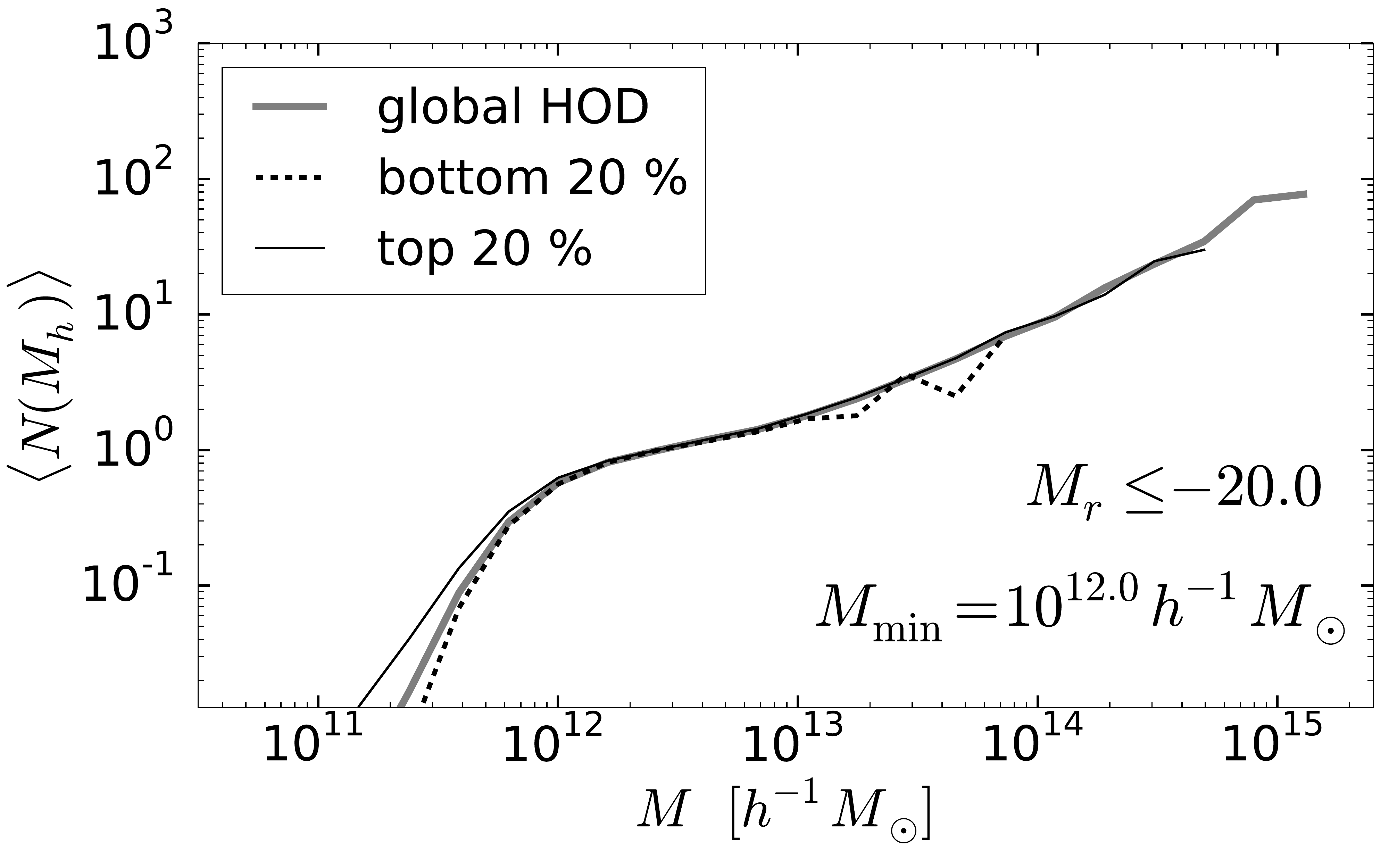}
 }
 
	\subfloat[]{
   \includegraphics[width=8cm]{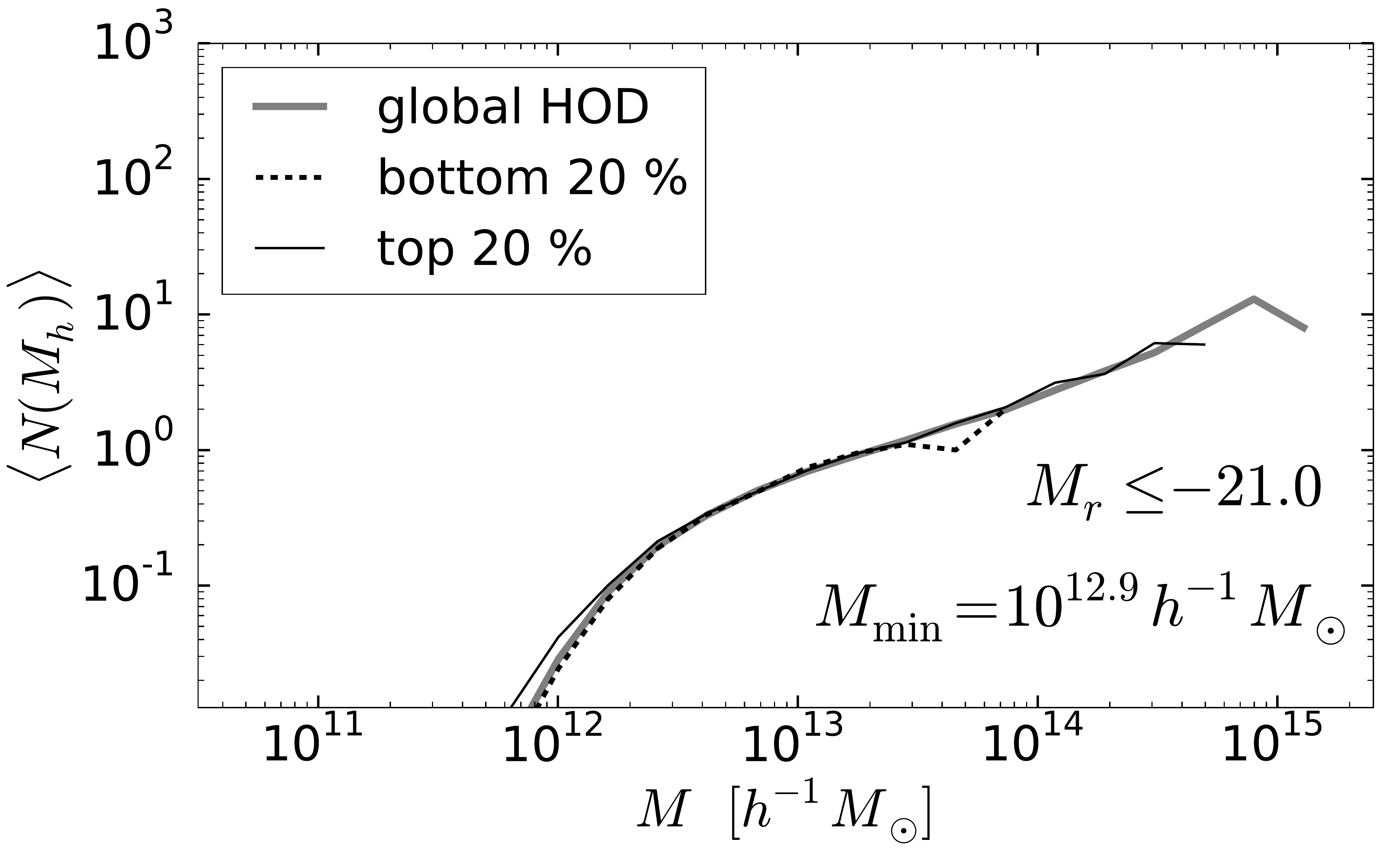}
}

	\subfloat[]{
   \includegraphics[width=8cm]{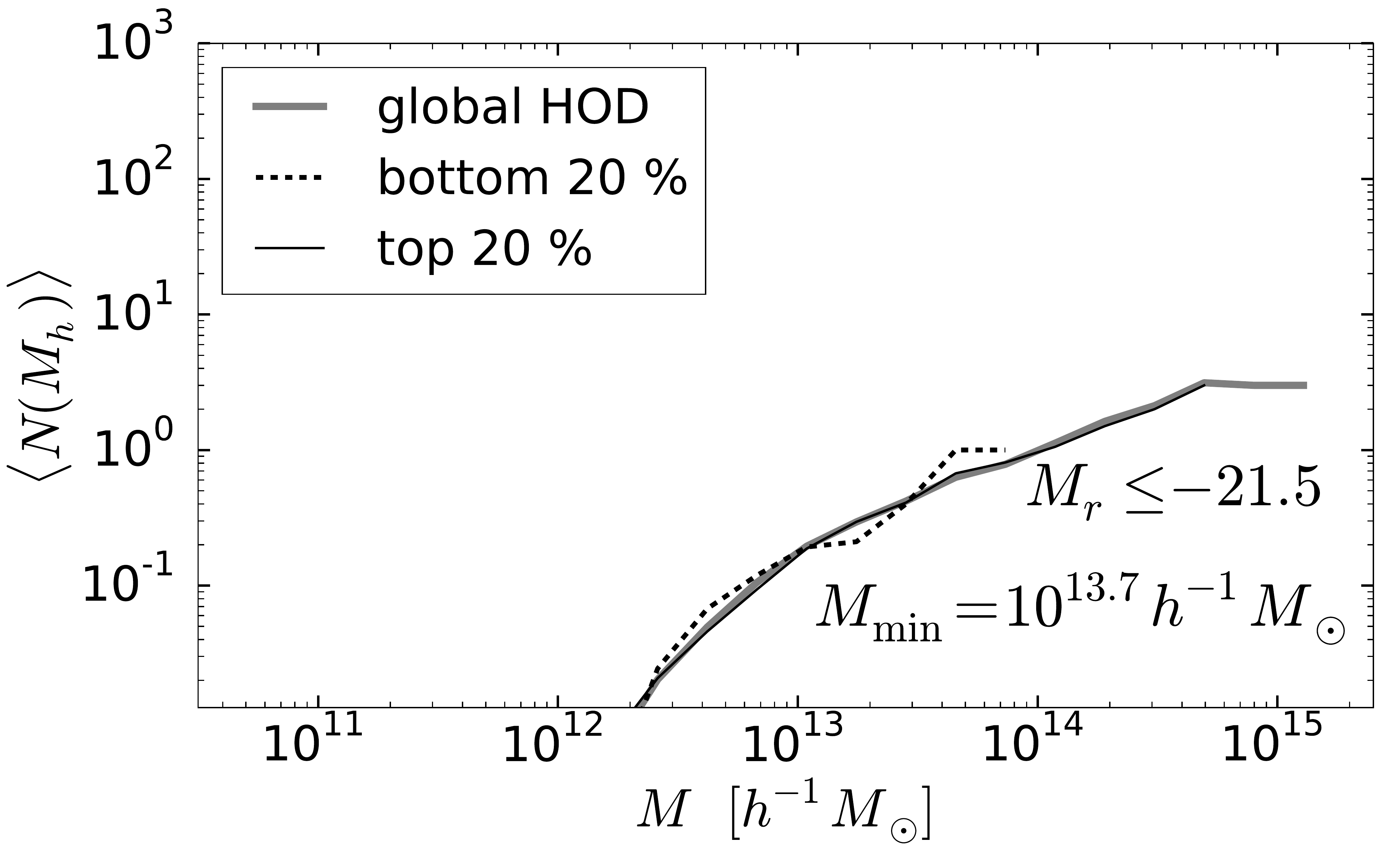}
}
\caption{Mean occupation functions of the HW13 catalogs for $M_r \leq -20$ (\textit{top}),  $-21$ (\textit{middle}), $-21.5$ (\textit{bottom}) for all halos and for halos in the $20 \%$ highest or lowest density environment measured by $\delta_{1-5}$, as in Fig. \ref{fig:19_HOD}. Galaxy assembly bias effects are smaller for more luminous samples.}
\label{fig:rest_HOD}
\end{figure}

Figure \ref{fig:rest_HOD} plots the measured HOD for HW13's $M_r \leq  -20$ (\textit{top}),  $M_r \leq  -21$ (\textit{middle}), and $M_r \leq  -21.5$ (\textit{bottom}) galaxy samples, comparing the global HOD to that of halos in the top 20th and bottom 20th percentile in environmental density, as in Fig. \ref{fig:19_HOD}.   Like the $M_r \leq -19 $ sample, the $M_r \leq -20$ sample shows an increase (decrease) in $\langle N \rangle$ for low mass host halos residing in higher (lower) density environments.   For brighter samples, the environmental dependence is weaker, and essentially indiscernible for $M_r \leq -21$ or  $M_r \leq-21.5$.  These results are consistent with the weakening impact of galaxy assembly bias at higher luminosities seen in Fig. \ref{fig:xi_bias}.  

\begin{figure}
\centering
\includegraphics[width=8cm]{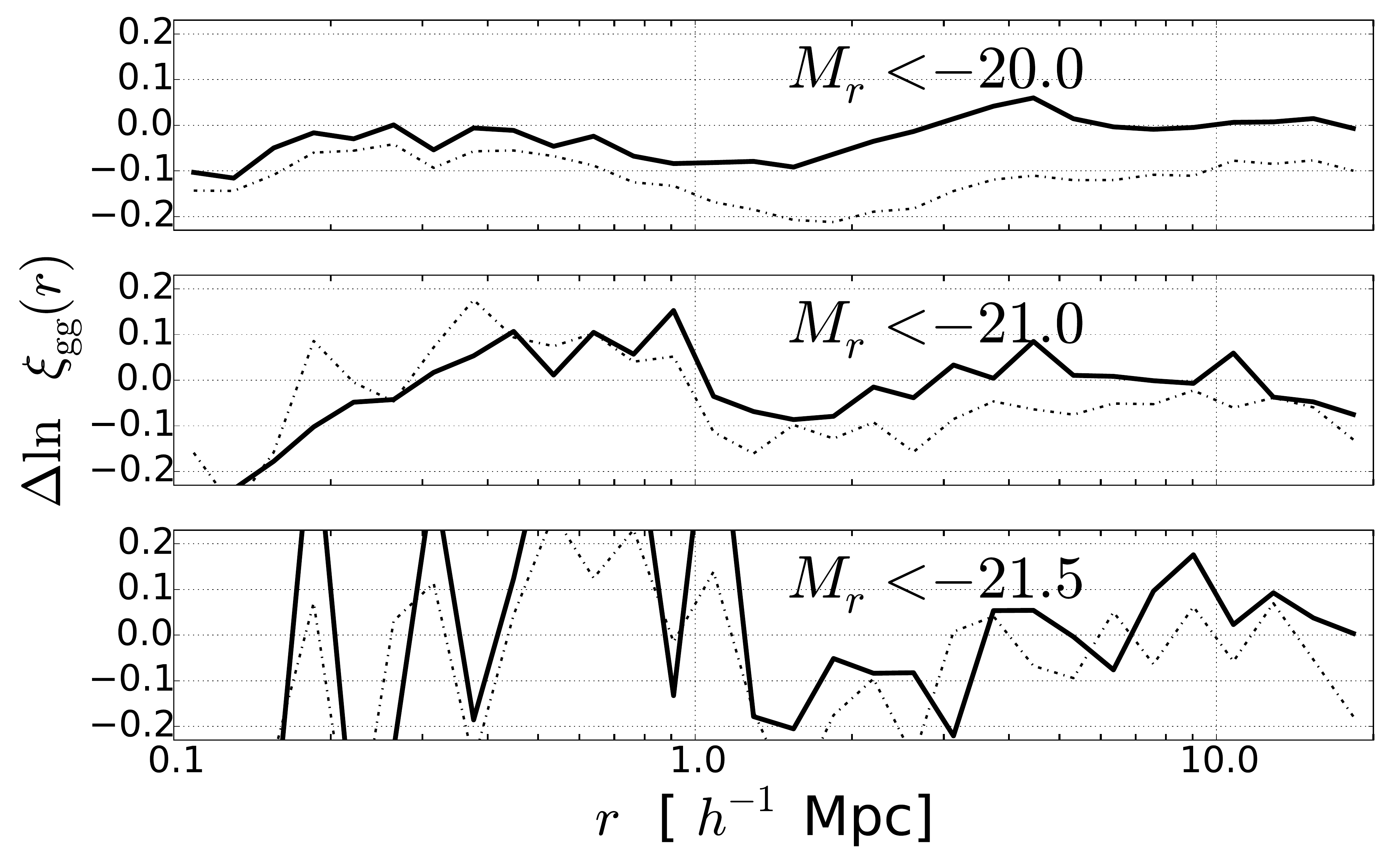}
\caption{Fractional deviations of $\xigg(r)$ from global (\textit{dashed}) and environmentally dependent (\textit{solid}) HOD catalogs compared to the HW13 catalogs for the $M_r \leq -20, -21, -21.5$ samples.  Similar to the bottom panel of Fig. \ref{fig:HOD_compare}.}  
\label{fig:HOD_xi_rest}
\end{figure}

\begin{figure}
\centering
\includegraphics[width=8cm]{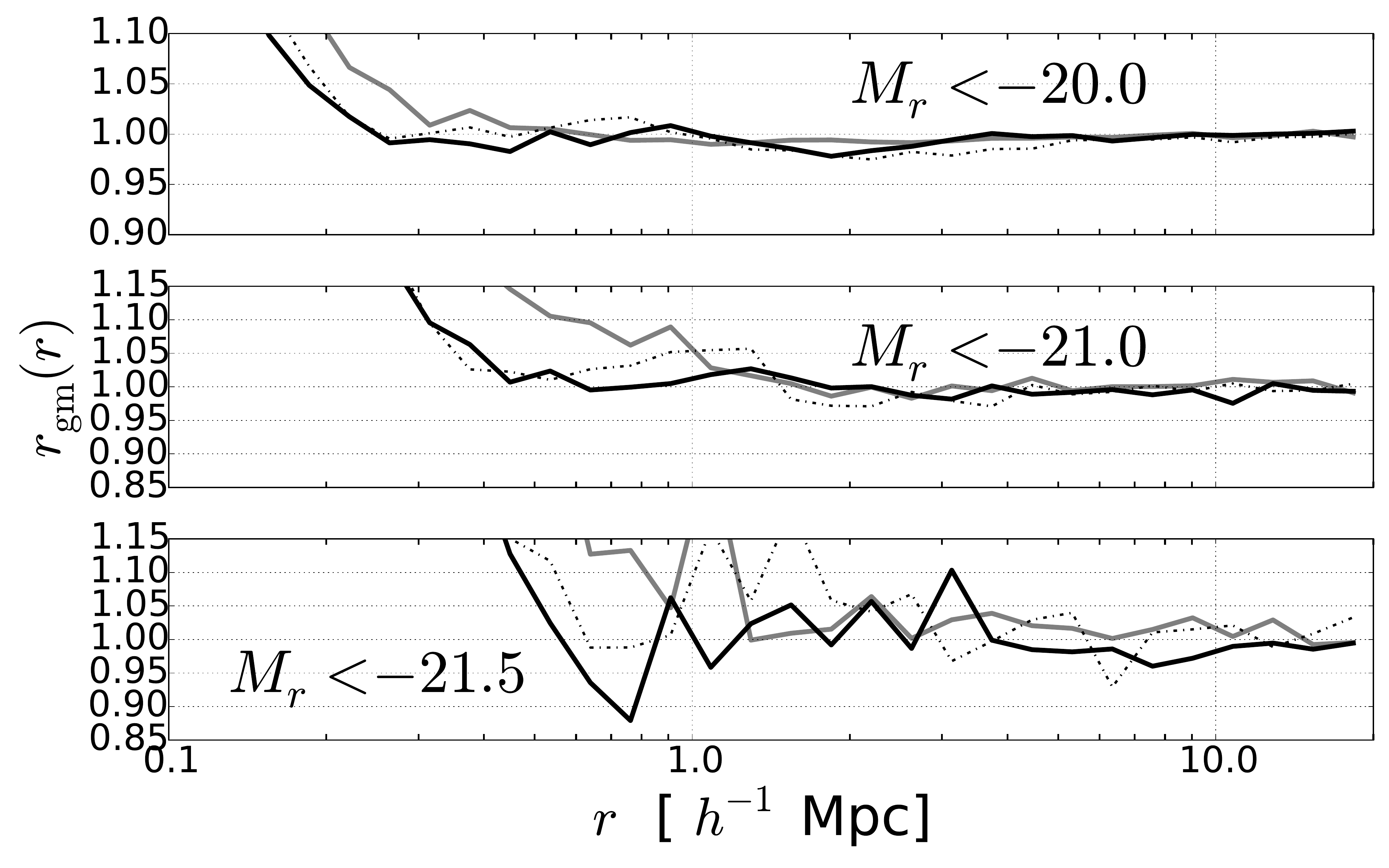}
\caption{Cross correlation coefficients $\rgm(r)$ from global (\textit{dashed}) and environmentally dependent (\textit{solid}) HOD catalogs compared to the HW13 catalogs (\textit{grey}) for the $M_r \leq -20, -21, -21.5$ samples.  Similar to Fig. \ref{fig:19_r_true}.}  
\label{fig:r_rest}
\end{figure}

Figure \ref{fig:HOD_xi_rest}, analogous to the lower panel of Fig. \ref{fig:HOD_compare}, plots the fractional difference in $\xigg(r)$ measured from the HW13 catalog and from catalogs constructed using the global HOD or the environmentally dependent HOD.  Results for $M_r \leq -20$ are similar to those for $M_r \leq -19$:  incorporating environmental dependence removes the $\sim 10 \%$ offset in the large scale bias factor found for the global HOD, but there are still 5-10 \% differences in $\xigg(r)$ for $r \approx 0.5 - 5 \hMpc$.  For $M_r \leq -21$ there is only a small bias offset for the global HOD model, and deviations in $\xigg(r)$ for the EDHOD model are consistent with random fluctuations.  For $M_r \leq -21.5$, all three models give consistent results.  Fig. \ref{fig:r_rest} shows results for the EDHOD and HOD cross-correlation coefficient compared to HW13 for our brighter samples.  For $M_r \leq -20$ and $-21$, EDHOD and HOD $\rgm(r)$ results track the HW13 results well on scales greater than $\sim 1 \hMpc$.  Results for $M_r \leq -21.5$ are dominated by noise. 

\begin{figure}
\centering
\includegraphics[width=8cm]{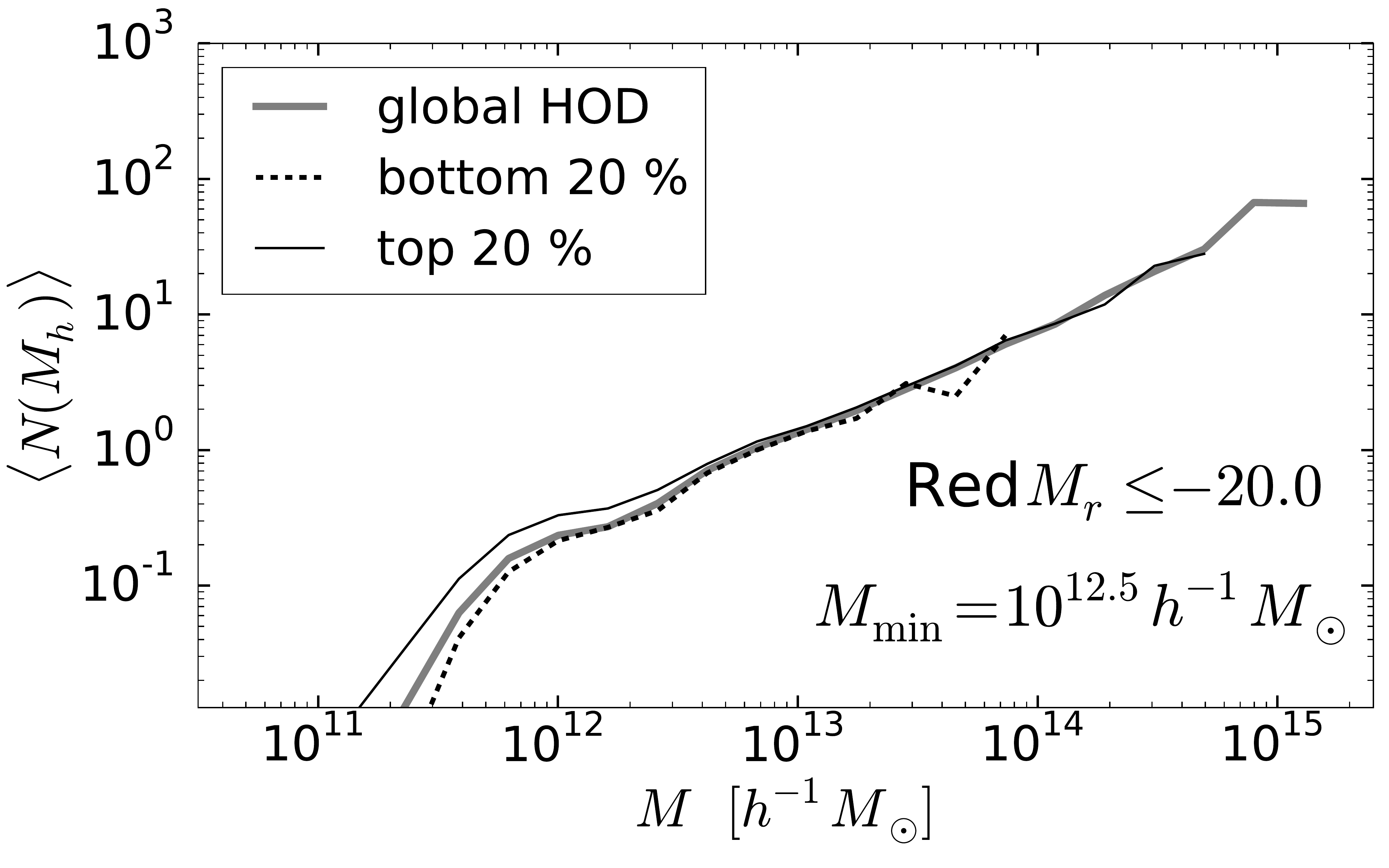}
\caption{Mean occupation for red galaxies with $M_r \leq -20$ in the HW13 catalog, in the same format as Figs. \ref{fig:19_HOD} and \ref{fig:rest_HOD}. }
\label{fig:20_HOD_red}
\end{figure}

Color selection has the potential to introduce stronger galaxy assembly bias because of the direct connection that the HW13 age-matching prescription introduces between colour and halo formation time.  Figure \ref{fig:20_HOD_red} shows the HOD environmental variation for the red $M_r \leq -20$ galaxies in the HW13 catalog.    Comparison to Fig. \ref{fig:rest_HOD}  shows that environmental dependence is indeed stronger than that of the full $M_r \leq -20$ sample; in particular, the increased $\langle N(M_h) \rangle$ in dense environments continues up to $10^{12.5} \hMsun$ halos.  However, there is still no indication of an environmental dependence of the satellite HOD.  

\begin{figure}
\centering
\includegraphics[width=8cm]{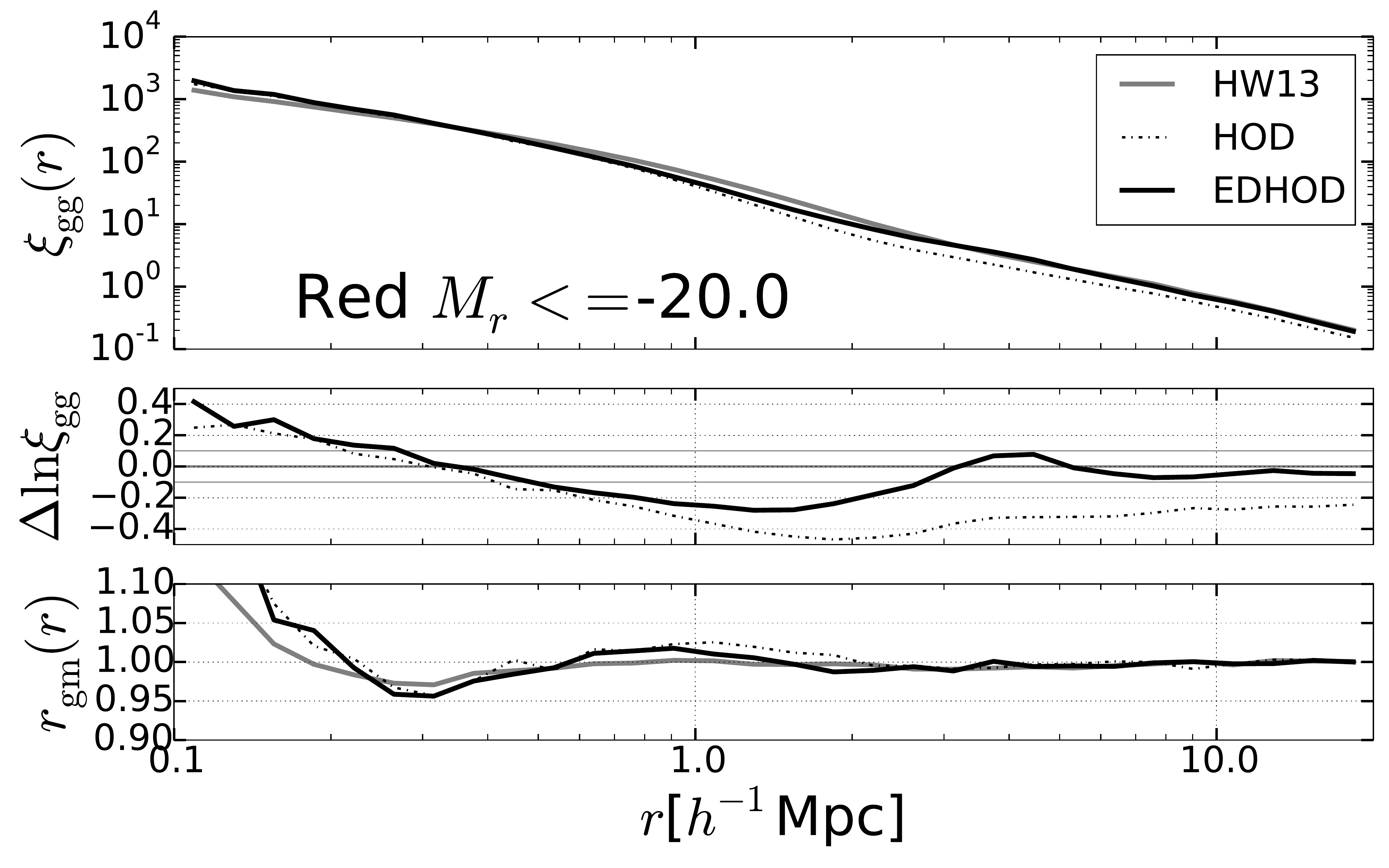}
\caption{Comparison of the galaxy correlation functions (\textit{top and middle}) and the galaxy-matter cross correlation coefficient (\textit{bottom}) for red $M_r \leq -20$ galaxies computed from the HW13 catalog (\textit{grey}) and catalogs created using the global (\textit{dashed}) or environmentally dependent (\textit{dot-dashed}) HODs.  Compare to Figs. \ref{fig:HOD_compare}, \ref{fig:19_r_true}, \ref{fig:HOD_xi_rest}. }
\label{fig:xi_red_20}
\end{figure}

Figure \ref{fig:xi_red_20} shows the deviations in the galaxy correlation function and galaxy-mass cross correlation coefficient for our red $M_r \leq -20$ galaxy sample.  As expected, differences between the global HOD model and the HW13 catalog are larger than those for the full $M_r \leq -20$ sample, with a 20 \% difference in the large scale bias factor.  The EDHOD model again removes this large scale offset but leaves significant deviations in the $0.5-5 \hMpc$ range.  Crucially, however, the values of $\rgm(r)$ computed from these three catalogs still match, at the $2 \%$ level or better for $r \geq 1 \hMpc$.

\subsection{Summary}

As shown previously by \cite{2014MNRAS.443.3044Z}, the HW13 galaxy catalogs exhibit substantial impact of assembly bias on galaxy clustering, particularly for low luminosity or colour selected samples.  In the case of luminosity threshold samples, galaxy assembly bias arises because HW13 tie luminosity to halo $V_{\rm max}$, which is correlated with formation time at fixed $M_h$.  For colour selected samples the connection to halo assembly is imposed directly by HW13's age matching procedure. 

In HOD terminology, the assembly bias manifests itself as a increase in $\langle N(M_h) \rangle$ for central galaxies of halos in denser than average environments, and a corresponding decrease of $\langle N(M_h) \rangle$  in low density environments.  We find no evident effect of halo environment on the satellite galaxy occupation.  Constructing HOD mock catalogs that incorporate the environmental dependence measured in the HW13 catalog removes the large scale offset in $\xigg(r)$ that arises with an environment-independent HOD model.  However, deviations of $\xigg(r)$ at the 5-20 \% level (depending on scale and galaxy sample) remain between the HW13 catalogs and catalogs constructed from an EDHOD.  Nonetheless, the HW13 catalogs, EDHOD catalogs, and (to a lesser extent) HOD catalogs yield similar predictions for $\rgm(r)$ at scales $r > 1\; \hMpc$, typically within the statistical fluctuations arising from the finite size of the simulated catalogs.  This similarity suggests that the impact of galaxy assembly bias on $\xigg(r)$ and $\xigm(r)$ will cancel in cosmological analysis, a point we address more directly in the next section.  

\section{Matter Clustering Inference}
In an observational analysis, one does not know the HOD or EDHOD of a galaxy sample {\it a priori} but infers it by fitting the observed galaxy clustering.  In a joint GGL + clustering analysis, the goal is to simultaneously infer the values of the cosmological parameters that determine $\Delta\Sigma(R)$.  A complete version of such an analysis would likely involve forward modeling of the projected clustering and GGL observables, with details that depend on the data sets being analyzed and on the external constraints adopted on the cosmological parameters (e.g., from CMB measurements).  Here we consider an idealized analysis in which de-projection has been used to translate $w_p(R)$ and $\Delta\Sigma(R)$ into the 3-d quantities $\xigg(r)$ and $\Omega_m\ximm(r)$.  HOD or EDHOD parameters are inferred by fitting $\xigg(r)$, and equation~(\ref{mass_inferred}) is used to infer $\Omega_m^2\ximm(r)$ from the GGL measurement.  We want to know whether this approach would yield unbiased estimates of $\rgm(r)$, and thus of $\Omega_m^2\ximm(r)$, given the galaxy assembly bias present in the HW13 abundance matching model. Our approach is inherently numerical, as we are using populated N-body halos to calculate $\xigg(r)$, $\xigm(r)$ and $\rgm(r)$ on all scales

\begin{figure}
\centering
\includegraphics[width=8cm]{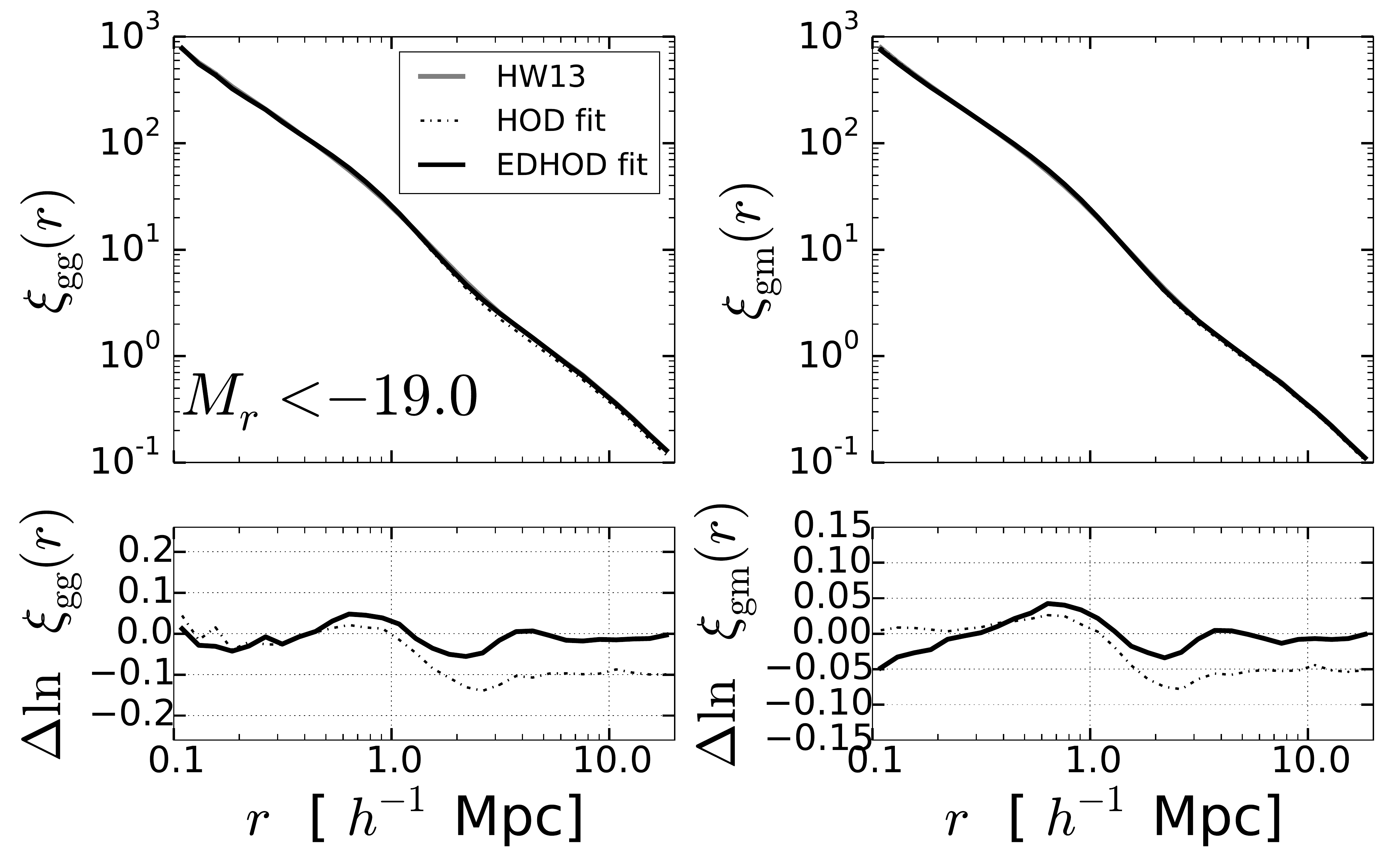}
\caption{HOD and EDHOD fitting results for the $M_r \leq -19$ sample.  Left and right panels show $\xigg(r)$ and $\xigm(r)$, respectively.  (ED)HOD parameters are inferred by fitting to $\xigg(r)$ over the range $0.1-30 \hMpc$ and including the total number of galaxies in HW13 as an additional fitting point. Lower panels show fractional deviations of the best-fit (ED)HOD models from the HW13 correlation functions. }
\label{fig:19_xi}
\end{figure}

\begin{figure}
\centering
\includegraphics[width=8cm]{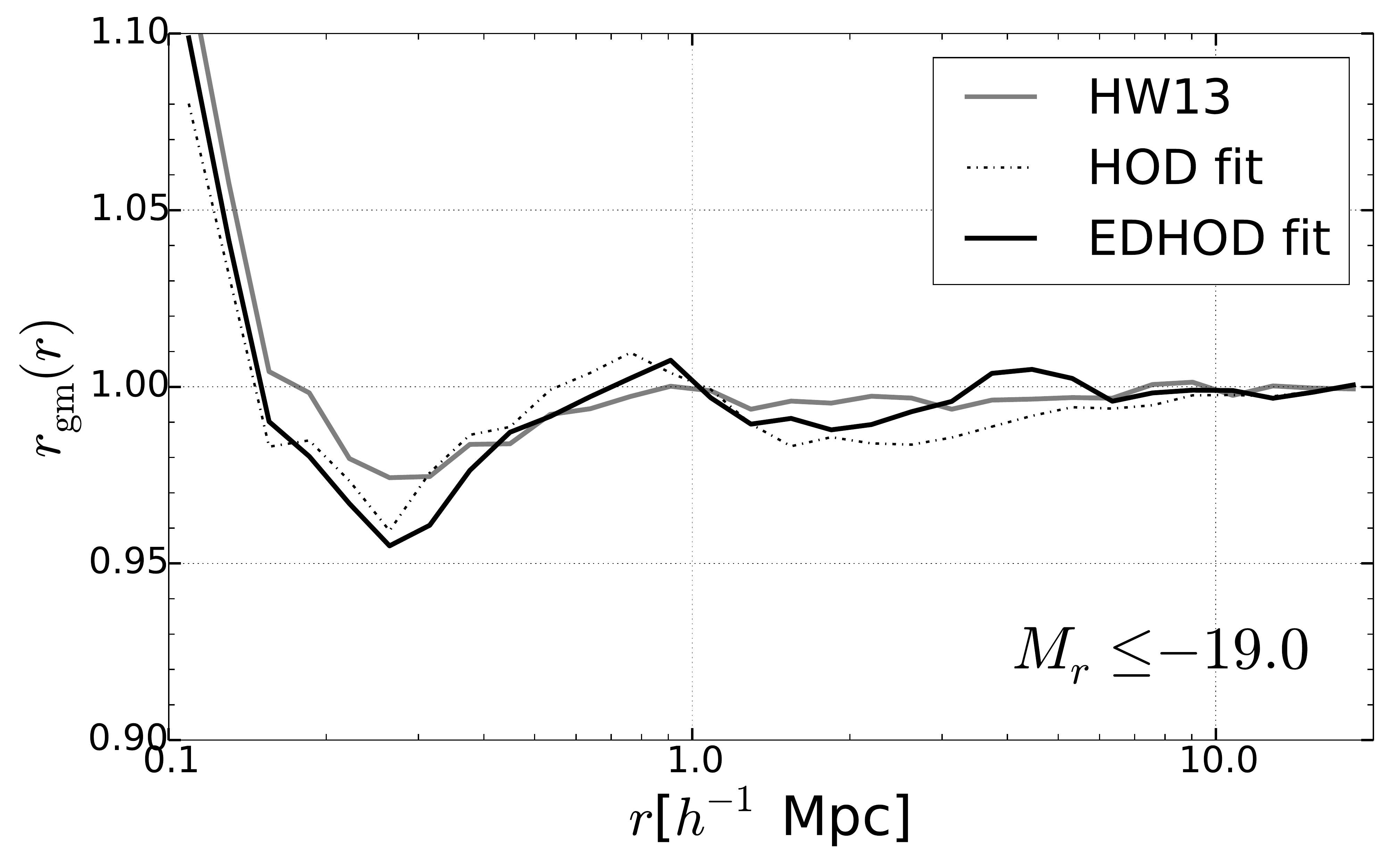}
\caption{Galaxy-matter cross correlation coefficient (eq. \ref{cross_corr}). for the $M_r \leq -19$ HW13 catalog (\textit{thick grey}) and for the EDHOD model (\textit{solid black}) and HOD (\textit{dot-dashed}) models that best fit the HW13 $\xigg(r)$ as shown in the left panels of Fig. \ref{fig:19_xi}.}
\label{fig:19_r}
\end{figure}

\begin{figure}
\captionsetup[subfigure]{labelformat=empty}
\centering
	\subfloat[]{
   \includegraphics[width=8cm]{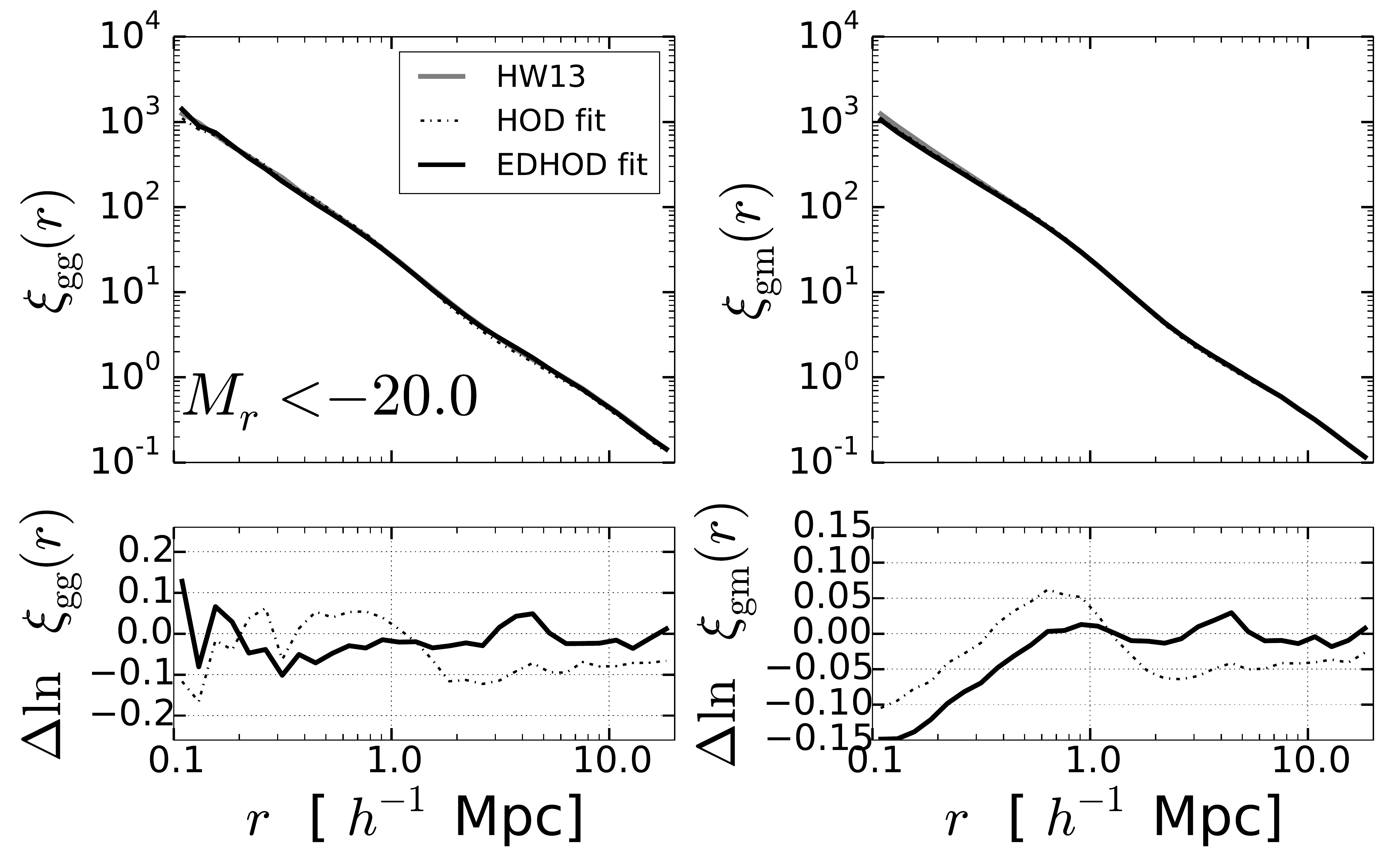}
}

	\subfloat[]{
   \includegraphics[width=8cm]{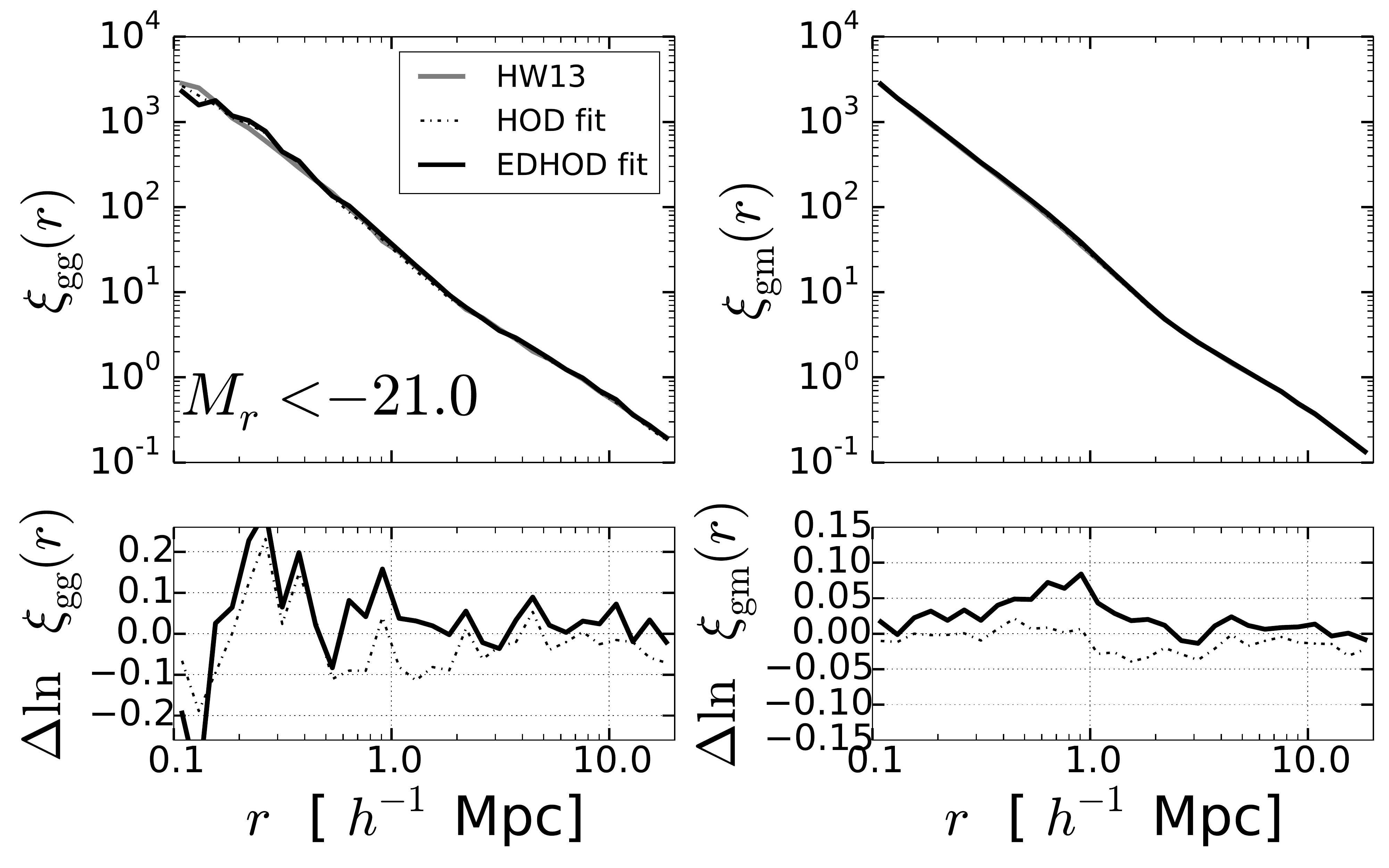}
}

	\subfloat[]{
   \includegraphics[width=8cm]{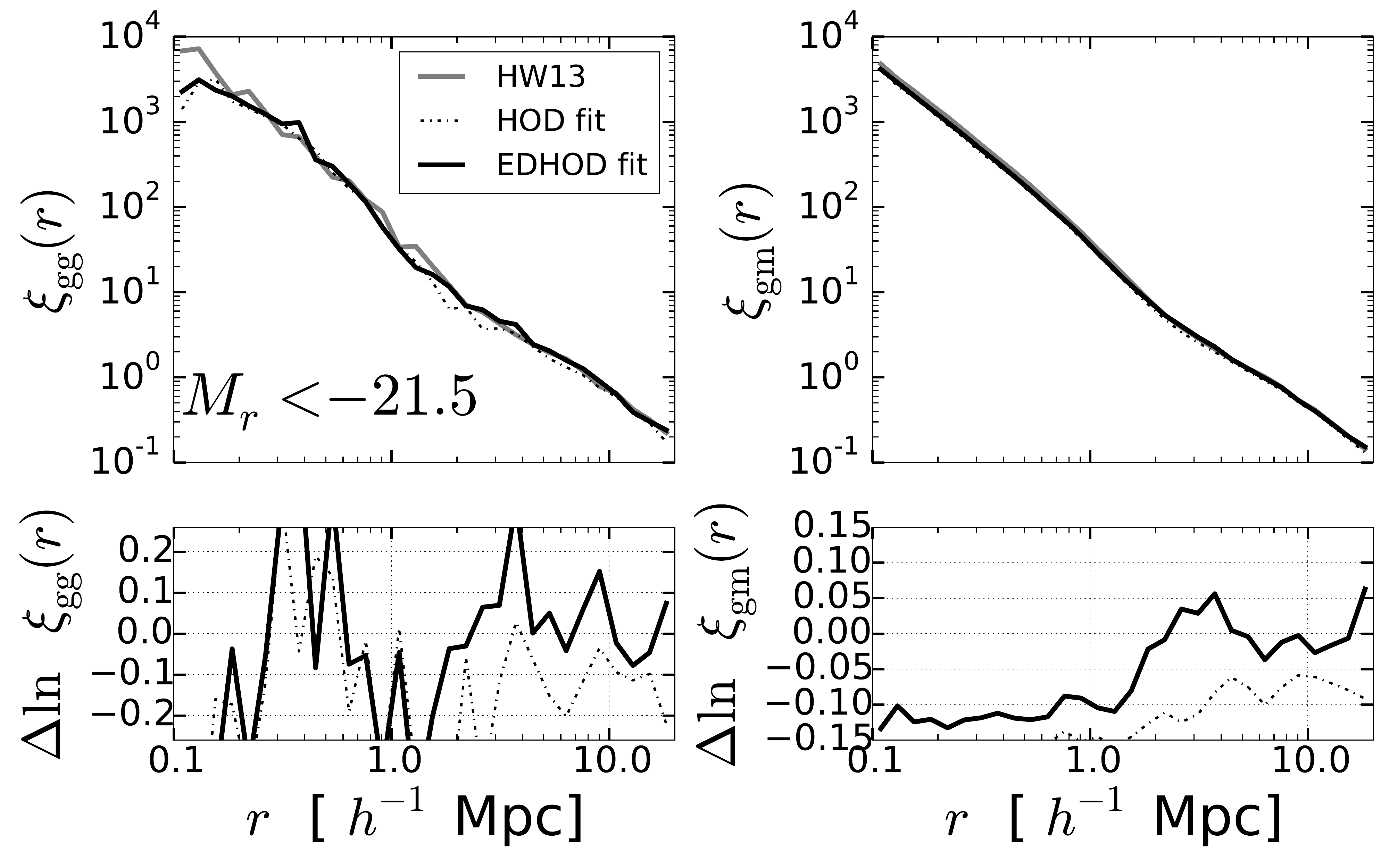}
}
\caption{Galaxy correlation function fitting results for the $M_r  \leq -20, -21, -21.5$ samples in the same format as in Fig. \ref{fig:19_xi}. }
\label{fig:rest_xi}
\end{figure}

\begin{figure}
\captionsetup[subfigure]{labelformat=empty}
\centering
	\subfloat[]{
   \includegraphics[width=8cm]{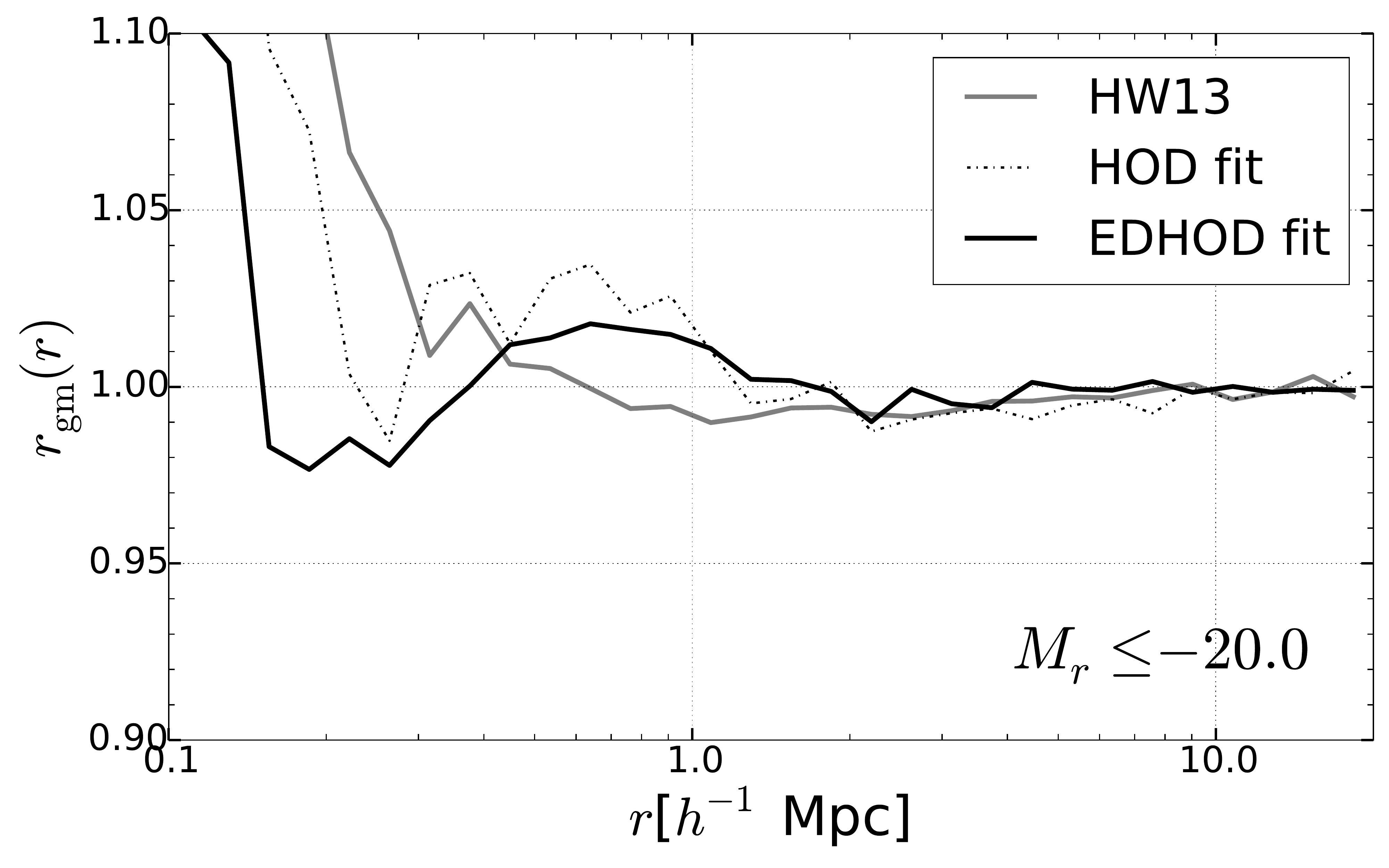}
}

	\subfloat[]{
   \includegraphics[width=8cm]{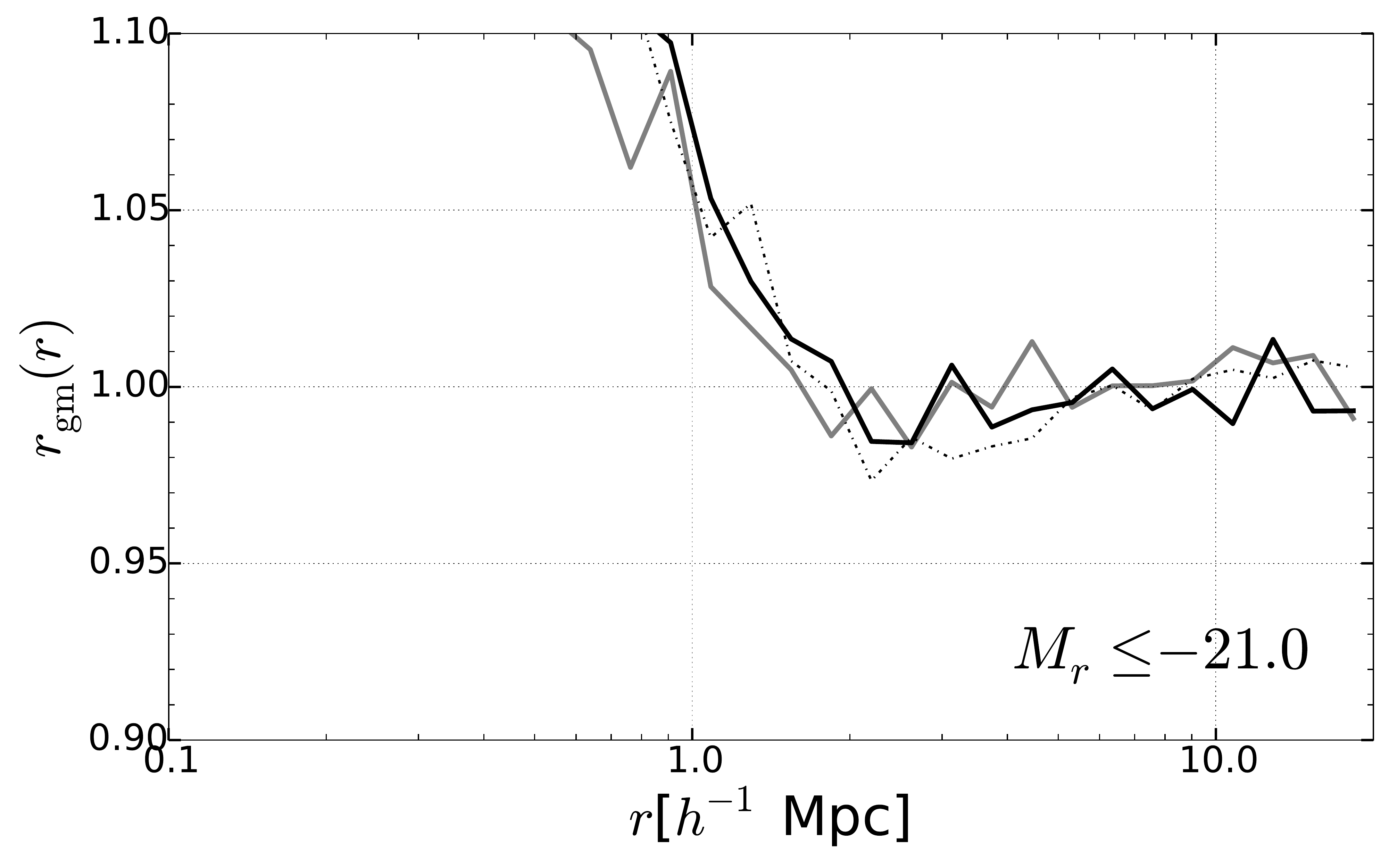}
}

	\subfloat[]{
   \includegraphics[width=8cm]{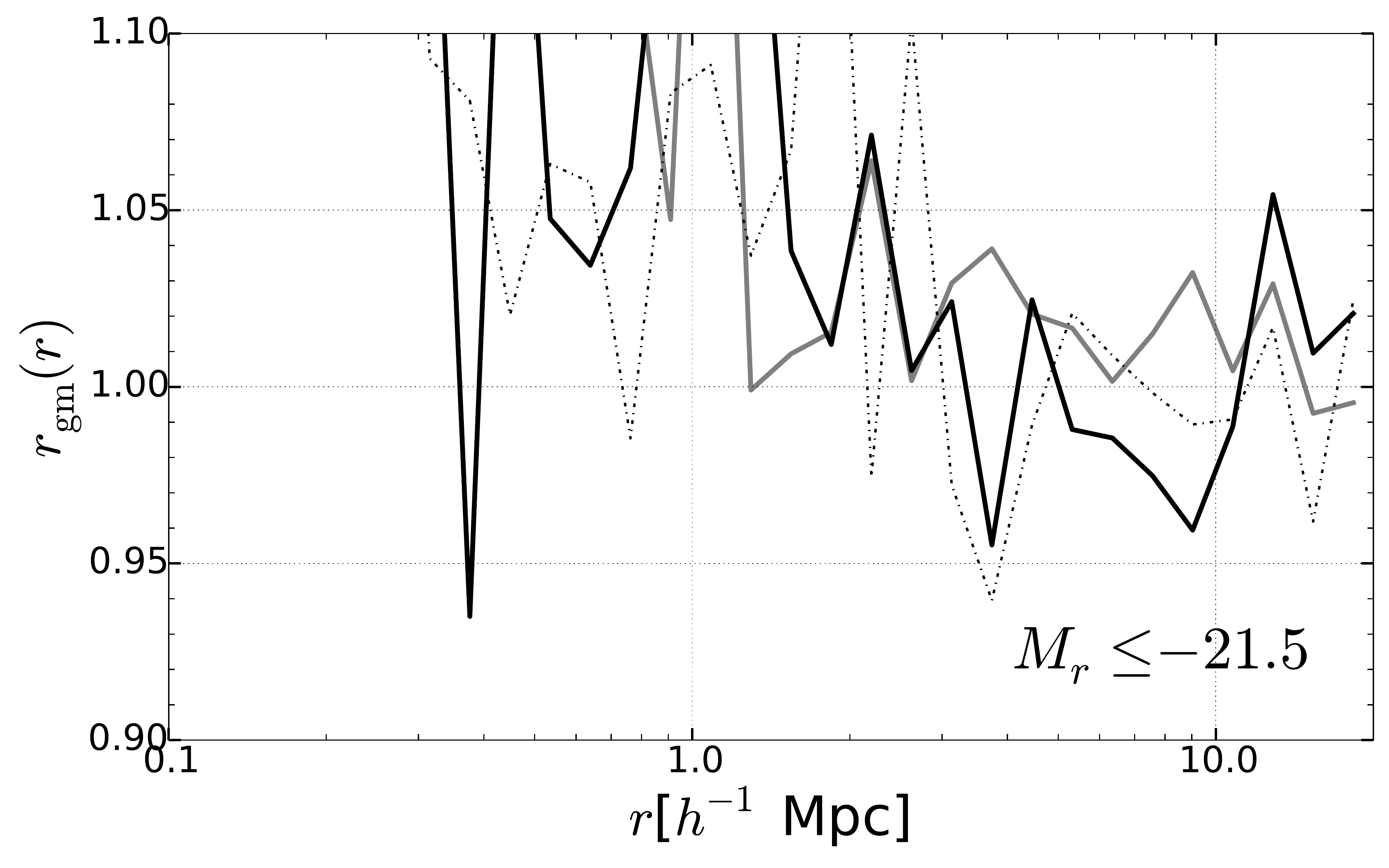}
}
\caption{Galaxy-matter cross correlation coefficients for the HW13 catalogs and the EDHOD and HOD model fit to the HW13 $\xigg(r)$, in the same format as Fig. \ref{fig:19_r}.}
\label{fig:rest_r}
\end{figure}

We construct an EDHOD model by allowing the central galaxy HOD parameters $M_\text{min}$ and $\sigma_{ \log M_h}$ to have a power-law dependence on $\delta_{1-5}$, in equations:
\begin{equation} \label{E_M} \log M_\text{min} = A + \gamma \log( \delta_{1-5}) ~,
\end{equation} 
\begin{equation} \label{E_S} \log \sigma_{\log M_h}= B + \beta \log (\delta_{1-5} ) ~.
 \end{equation} 
Note that our EDHOD model has seven parameters, $\{A, B, \gamma, \beta, M_1, M_{\rm cut}, \alpha\} $, whereas the HOD presented in $\S 2$ has five.  (ED)HOD parameters are inferred by  using a downhill simplex method to minimize a sum of squares function, $\sum_i (D^i_\text{model} - D^i_\text{HW13})^2/(D^i_\text{HW13})^2$.  The data vector $\vec{D}$ is  galaxy-correlation function in the range $r=0.1-30 \; \hMpc$ in $30$ equal logarithmic bins, with the total number of galaxies included as an additional fitting point for each sample selection.

Figure \ref{fig:19_xi} shows the results of fitting the $M_r \leq -19$ galaxy sample.  The EDHOD model achieves a good overall fit to the HW13 $\xigg(r)$, with fluctuating deviations up to $\sim 5 \%$ in the region of the 1-halo to 2-halo crossover.  The best-fit HOD model has a large scale offset of $10\%$ in $\xigg(r)$ (5 \% in $b_g$); while several HOD parameters can be adjusted to increase the large scale bias, doing so would alter the small scale $\xigg(r)$ in a way that worsens the overall fit.  This kind of offset could be a diagnostic for galaxy assembly bias, but we have not explored whether it can be erased by giving more freedom to the assumed radial profile of satellites within halos.  The right panel of Fig. \ref{fig:19_xi} shows $\xigm(r)$ predicted by the HOD or EDHOD model that best fits $\xigg(r)$.  Deviations from the HW13 $\xigm(r)$ are similar to those for $\xigg(r)$, but reduced in magnitude by a factor of two. 

Figure \ref{fig:19_r} compares the cross-correlation coefficients for $M_r \leq -19$ samples in the HW13, HOD, and EDHOD catalogs.  In each case we use the simulation's true $\ximm(r)$ and the $\xigg(r)$ and $\xigm(r)$ computed numerically from the corresponding catalog, calculating $\rgm(r)$ from Eqn. \ref{cross_corr}.  At $r > 1 \; \hMpc$, the EDHOD and HOD models reproduce the HW13 result to 1 \% or better, despite the 5-10 \% deviations in $\xigg(r)$.  For this sample, all three models predict $\rgm(r)$ very close to one on these scales.  Within the 1-halo regime, the EDHOD and HOD fits continue to track the HW13 result, with deviations of a few percent.  

Figure \ref{fig:rest_xi} shows similar results for the $M_r \leq -20, -21, \; \text{and} \; -21.5$ samples.  Results in $\S$ 2 show that the impact of galaxy assembly bias decreases with increasing luminosity threshold.  Consistent with this behavior, the EDHOD model fits the HW13 $\xigg(r)$ substantially better than the global HOD model, but the difference between EDHOD and HOD fits becomes less significant for the higher thresholds.  Results for these brighter, sparser samples become progressively noisier.   As seen previously for $M_r \leq -19$, deviations in $\xigg(r)$ are mirrored in $\xigm(r)$, with a factor $\sim 2$ reduction in amplitude. 

Figure \ref{fig:rest_r} shows the cross-correlation coefficients for these higher luminosity samples.  For $M_r \leq -20$ and $M_r \leq -21$, the EDHOD and HOD models again track the HW13 results at $r > 1\; \hMpc$, with 1-2 \% deviations in $\rgm(r)$ that appear consistent with random fluctuations.  In the $M_r \leq -21$ case, this percent-level agreement holds even when $\rgm(r)$ has climbed to 1.05, so it is not simply a consequence of all three models predicting $\rgm \approx 1.0$.  For the $M_r \leq -21.5$ samples, results are too noisy for percent-level tests, but there is no evidence for a systematic difference between the HW13 cross-correlation and those predicted by the HOD or EDHOD fits. 

Given measurements of $\xigg(r)$ and $\xigm(r)$, and a cross correlation coefficient $\rgm(r)$ inferred from an HOD or EDHOD model fit, one can calculate the underlying matter correlation function $\ximm(r)$ via Eqn. \ref{mass_inferred}.  (Because GGL constrains $\Omegam \xigm$, it is $\Omegam^2 \ximm$ that is constrained, but here we omit the $\Omega_\text{m}$ dependence for simplicity.)  

\begin{figure*}
\centering
\includegraphics[width=16cm]{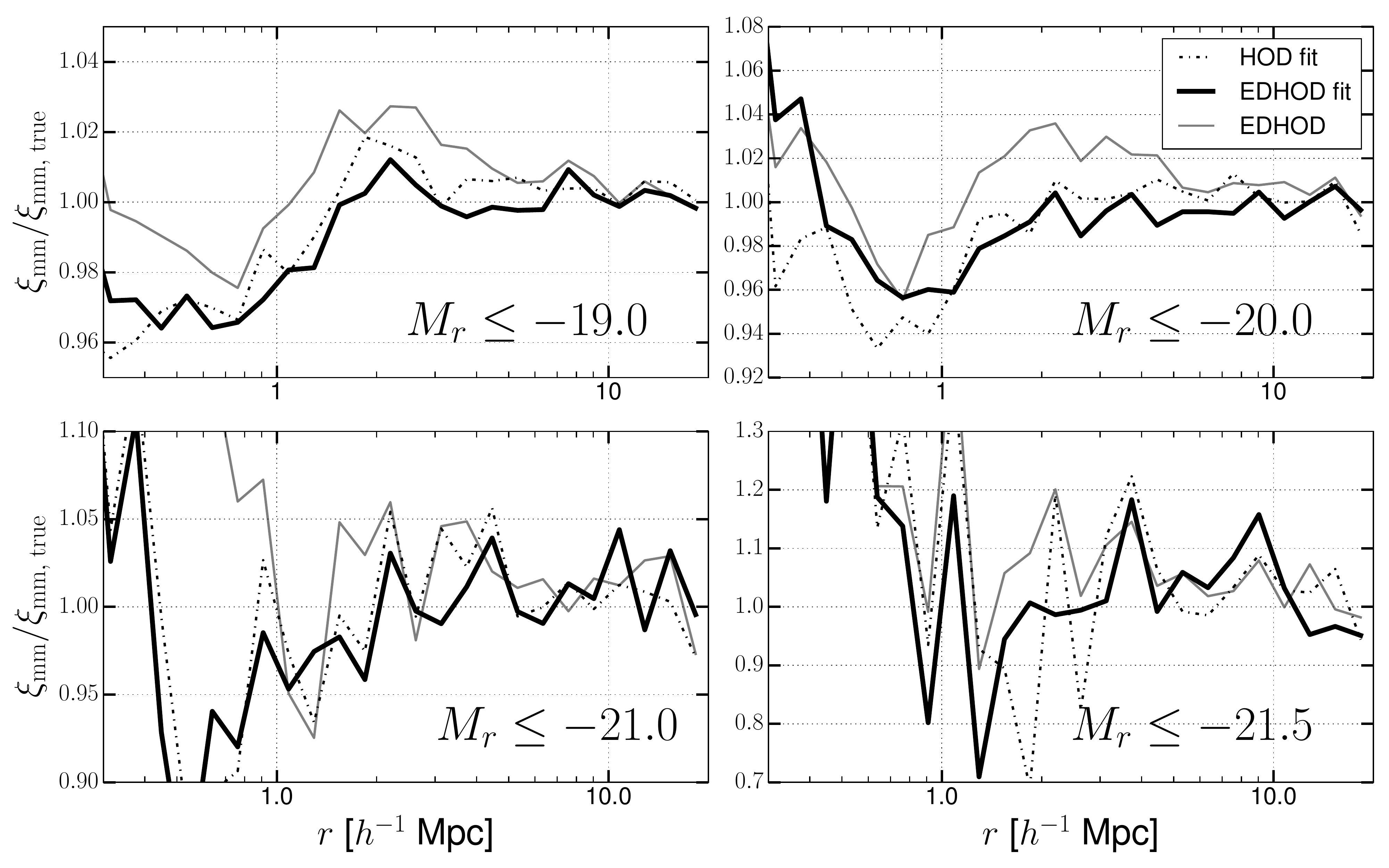}
\caption{Accuracy of the matter correlation functions inferred from the $\xigg (r)$ and $\xigm (r)$ measurements of the four HW13 catalogs, using Eqn. \ref{mass_inferred} with $\rgm (r)$ computed from the EDHOD (\textit{solid}) or HOD (\textit{dot-dashed}) model to fit to $\xigg (r)$.  Each panel plots the ratio of the recovered $\ximm(r)$ to the true $\ximm (r)$ measured in the Bolshoi simulation.  Grey lines show the effect of using the EDHOD directly measured from the HW13 catalogs instead of that inferred by fitting $\xigg (r)$. }
\label{fig:mass_inferred_fit}
\end{figure*}

Solid black curves in Fig. \ref{fig:mass_inferred_fit} show the principal results of this paper, the accuracy within which the true matter correlation function of the Bolshoi simulation is recovered by applying this procedure to the $\xigg(r)$ and $\xigm(r)$ measurements from the HW13 catalogs, using an EDHOD model to infer $\rgm(r)$.  For the $M_r \leq -19$ samples, recovery is accurate to better than $2 \%$ for $r > 1 \; \hMpc$.  Deviations for the brighter samples are larger, but they appear consistent with statistical fluctuations.  Deviations are larger inside $1 \; \hMpc$, but not drastically so.  Results for the global HOD fits are similar to those for the EDHOD fits, even though the global HOD model does not produce good fits to $\xigg(r)$.  Grey solid curves in Fig. \ref{fig:mass_inferred_fit} show the results of using the true EDHOD measured directly from the HW13 catalogs (as described in $\S 2$).  Fitting the EDHOD to $\xigg(r)$ yields better recovery of $\ximm(r)$ than using the directly measured EDHOD.  

\begin{figure}
\captionsetup[subfigure]{labelformat=empty}
\centering
\subfloat[]{
   \includegraphics[width=8cm]{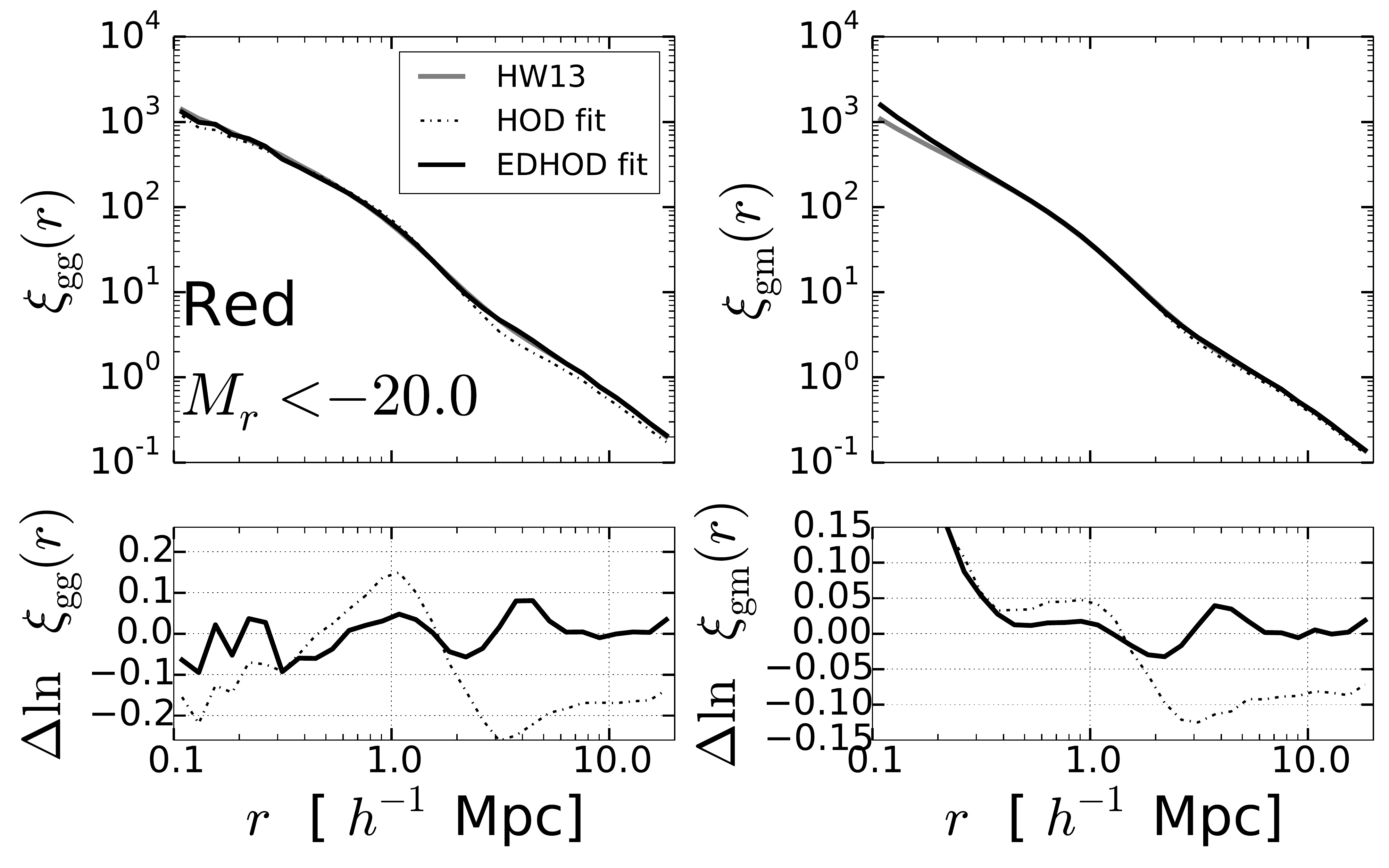}
}

\subfloat[]{
   \includegraphics[width=8cm]{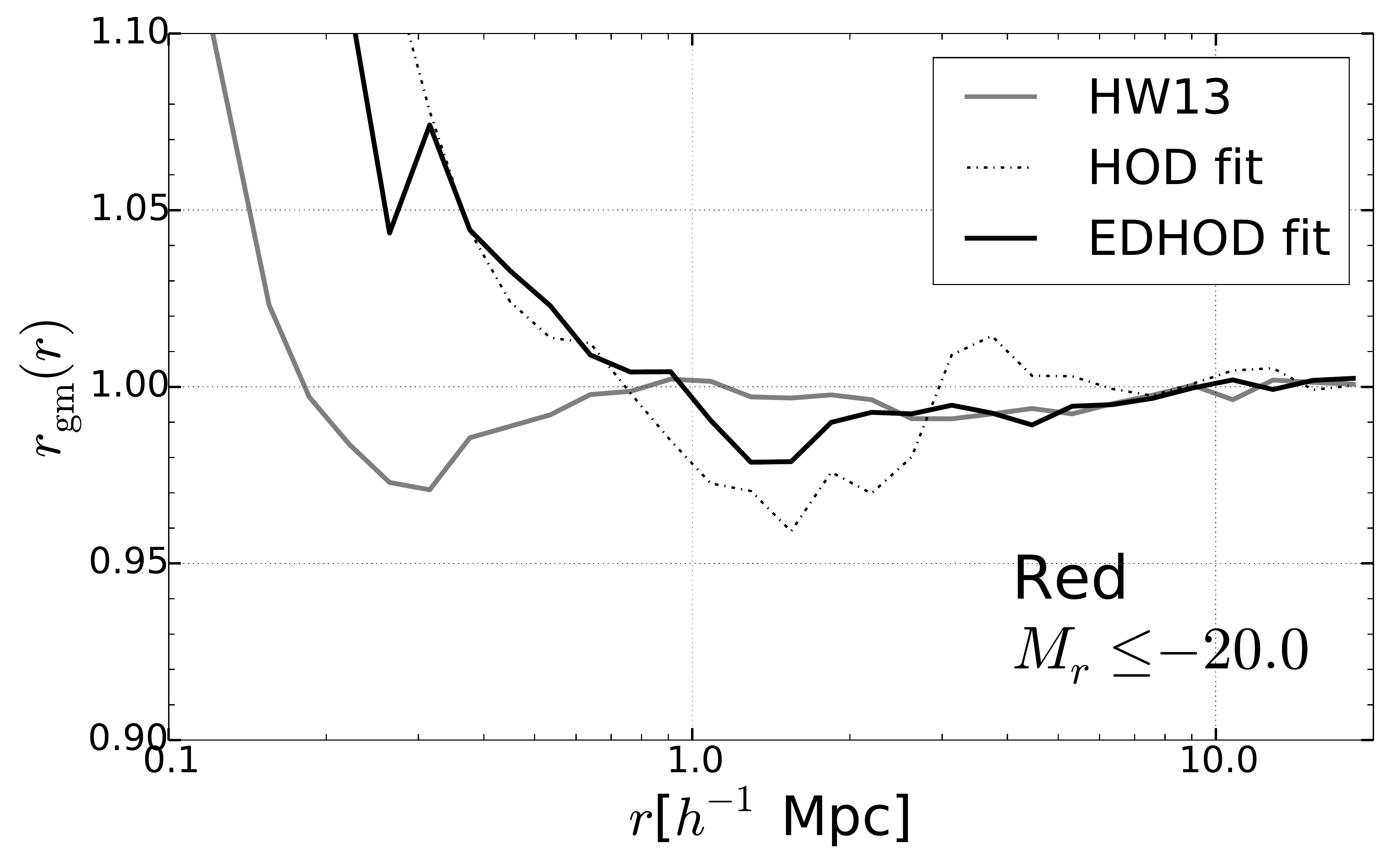}
}

\subfloat[]{
   \includegraphics[width=8cm]{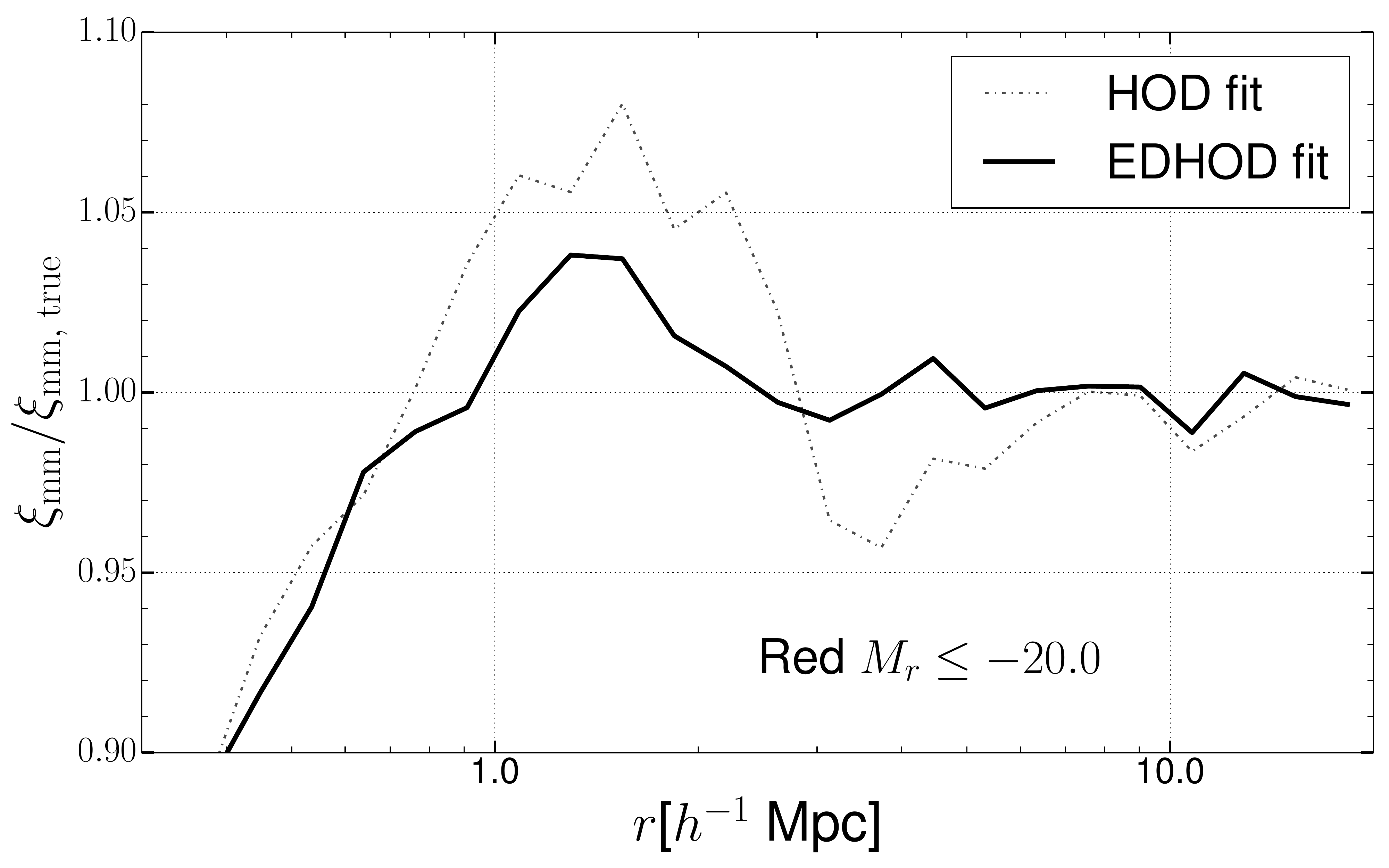}
}
\caption{Correlation function and matter correlation recovery for the red $M_r \leq -20$ galaxies.  Upper panels show the HOD and EDHOD fits to $\xigg(r)$ and the predicted $\xigm(r)$, in the format of Fig. \ref{fig:19_xi}.  The central panel shows $\rgm(r)$ for the HW13 catalog and the two HOD catalogs, as in Fig. \ref{fig:19_r}.  The bottom panel, analogous to Fig. \ref{fig:mass_inferred_fit}, shows $\ximm(r)$ inferred from the measured $\xigg(r)$ and $\xigm(r)$, using $\rgm(r)$ shown in the middle panel. }
\label{fig:red_analysis}
\end{figure}

A red galaxy sample presents a stringent test of our methodology because of the strong galaxy assembly bias imprinted by the HW13 age-matching procedure, and secondarily because the HOD paramerization that we use is designed for luminosity threshold samples.  Figure \ref{fig:red_analysis} shows results for the red $M_r \leq -20$ galaxy sample, similar to those shown in Figs. \ref{fig:19_xi}-\ref{fig:mass_inferred_fit} for the luminosity threshold samples.  For red galaxies, the global HOD model produces a poor fit to $\xigg(r)$, with a 10 \% large scale offset and a maximum deviation of 25 \%.  The EDHOD fit has deviations of 5-10 \% in the $1-10 \; \hMpc$ range.   Nonetheless, the EDHOD model matches the HW13 cross-correlation coefficient to 1 \% or better at $r > 2 \; \hMpc$, and to 2 \% or better at $r > 1 \; \hMpc$.  Even for the global HOD model, the agreement in $\rgm$ is generally better than 3 \% at $r > 1 \; \hMpc$.

The end result, shown in the bottom panel of Fig. \ref{fig:red_analysis}, is that our EDHOD modeling allows recovery of $\ximm(r)$ with accuracy of 2 \% or better at $ r > 2 \; \hMpc$ when applied to the $M_r \leq -20$ red galaxy population of the HW13 catalog.  Even though this model does not fully represent the galaxy assembly bias present in the HW13 catalog, errors in $\xigg(r)$ and $\xigm(r)$ cancel in a way that accurately estimates matter clustering.  In contrast to the luminosity-threshold cases shown in Fig. \ref{fig:mass_inferred_fit}, the EDHOD modeling significantly outperforms the global HOD modeling in this case, so the additional complication appears worthwhile at least in situations where the global HOD model yields a poor fit to $\xigg(r)$.

\section{Conclusion}

The combination of galaxy clustering and GGL is a powerful complement to cosmic shear measurements of matter clustering, applicable to the same data sets but with statistical and systematic errors that are at least partly independent.  Fully exploiting the combination requires a model of galaxy correlations and galaxy-matter cross correlations that extends to the non-linear regime.  Halo occupation methods are a natural approach to this problem, as advocated by \cite{2006ApJ...652...26Y,2009MNRAS.394..929C,2012MNRAS.426..566C,2011ApJ...738...45L,2012PhRvD..86h3504Y,2013MNRAS.430..767C,2015MNRAS.449.1352C,2015ApJ...806....2M}.  However, modeling that assumes an environment independent HOD could yield a biased result if the galaxy population is significantly affected by assembly bias, which can alter the galaxy content of halos of fixed mass in different large scale environments.  In this paper, we have used the abundance matching and age-matching catalogs of \citealt{2013MNRAS.435.1313H} (HW13), which exhibit substantial galaxy assembly bias, to show that systematic error in the recovery of matter clustering remain small even down to scales of $\sim 1 \hMpc$. 

The HW13 catalogs ties its galaxy properties to halo assembly history by using $V_\text{max}$ (rather than halo mass) as a ranking parameter for luminosity and formation time as a ranking parameter for color.  As shown in  \cite{2014MNRAS.443.3044Z} and $\S 2 $ of this paper, the HW13 scheme thereby translates halo assembly bias into galaxy assembly bias.  This galaxy assembly bias has substantial (5-50 \%) impact on galaxy correlations and cross-correlations, with the largest effects for low luminosity or color-defined samples.  The HOD of central galaxies in these samples is boosted in dense environments (defined by overdensity in a $1-5\; \hMpc$ spherical annulus) and suppressed in low density environments.  We find no significant variation of the satellite galaxy HOD with environment for any of these samples.  By incorporating the environment dependence of the central galaxy HOD, we construct EDHOD catalogs that reproduce the large scale bias of the HW13 catalogs, but these still have 5-10 \% deviations of $\xigg(r)$ at scales of $1-5 \; \hMpc$.  Nonetheless, these catalogs predict nearly the same galaxy-matter cross-correlation coefficient $\rgm(r)$ as the corresponding HW13 catalog on scales $r > 1\; \hMpc$, typically to $< 2 \%$ or within the statistical fluctuations of the finite simulation volume.  Catalogs that use the global HOD with no environment dependence preform only slightly worse in reproducing $\rgm(r)$, even though they have substantial errors in $\xigg(r)$ and $\xigm(r)$ individually.

Our most important results come from treating the HW13 catalogs as a source of ``observed" $\xigg(r)$ and $\xigm(r)$, then attempting to recover the true matter correlation function $\ximm(r)$ measured directly from the Bolshoi simulation by fitting (ED)HOD parameters to $\xigg(r)$.  We consider luminosity threshold samples $M_r \leq -19, -20, -21, \; \text{and}\; -21.5$, and color-selected sample of red galaxies with $M_r \leq -20$.  We construct EDHOD catalogs that adopt power-law trends of the central galaxy $M_{\rm min}$ and $\sigma_{\log M_h}$ parameters with $\delta_{1-5}$ (Eqns. \ref{E_M} and \ref{E_S}), determining their parameters by fitting $\xigg(r)$ in the range $0.1 - 30 \; \hMpc$.  We also fit global HOD models with no environment dependence and construct corresponding catalogs. We calculate $\rgm(r)$ numerically from these catalogs and apply this function to the measured $\xigg(r)$ and $\xigm(r)$ to infer $\ximm(r)$, via Eqn. \ref{mass_inferred}.  While this procedure is idealized compared to a true observational analysis, it enables us to use numerical calculations instead of analytic approximations (which are not sufficiently accurate on these scales), and it isolates the key physical issue --- the influence (or lack of influence) of galaxy assembly bias on the cross-correlation coefficient.  

For dense galaxy samples that yield precise measurements, our procedure recovers $\ximm(r)$ to 2\% or better on scales $r > 1 \; \hMpc$ ($M_r \leq -19$), $1.5 \; \hMpc$ ($M_r \leq -20$), or $2 \; \hMpc$ ($M_r \leq -20$ red).  For sparser, high luminosity samples, recovery of $\ximm(r)$ appears consistent with statistical uncertainties at $r > 1\; \hMpc$.  Results using HOD or EDHOD are usually similar, but the EDHOD recovery is more accurate for the red galaxy sample, which has the strongest assembly bias. 

On linear scales, where galaxy bias is described by a single parameter $b_\text{g}$, the detailed physics that produces that bias does not matter for GGL + clustering analysis, a point exploited by the observational study of \cite{2013MNRAS.432.1544M}.  Our analysis of the HW13 catalogs, which exhibit fairly complex assembly bias because of the prescription used to create them, suggest that this insensitivity continues down to $\sim 1 \;\hMpc$ scales, even though bias becomes scale-dependent and $\rgm$ can deviate from unity in this regime.  

There are many interesting avenues for future investigations.  One is to apply HW13-like prescriptions to larger simulations and higher redshifts.  This would enable higher precision tests of $\ximm(r)$ recovery, especially for sparse samples of luminous galaxies like those in the SDSS Luminous Red Galaxy survey (\citealt{2001AJ....122.2267E}) or the SDSS-III BOSS sample (\citealt{2013AJ....145...10D}).  It would also be informative to ``stress-test" our findings against models with more extreme galaxy assembly bias or non-linear $\rgm$ values further from unity.  We would also like to apply these methods to catalogs created with semi-analytic galaxy formation models, and to hydrodynamic simulations with large enough volume to yield good statistics for $\xigg(r)$ and $\xigm(r)$.  In addition to exhibiting different (and perhaps weaker, cf. \cite{2014PhDT.......126M,2015arXiv150701948C}) galaxy assembly bias, hydrodynamic simulations are important for predicting the impact of gas physics, star formation, and feedback on the small scale mass distribution (e.g., \citealt{2014MNRAS.440.2997V}). 

For observational applications, significant work is needed to develop a forward modeling framework that directly predicts observables such as $w_{\rm p}(R)$ and $\Delta \Sigma(R)$.  A critical element of such a framework is a numerically calibrated procedure to predict $\xigg(r)$ and $\xigm(r)$ as a function of cosmological and EDHOD parameters, since existing analytic approximations have errors at the several percent level on scales of interest (e.g., \citealt{2006ApJ...652...26Y,2013MNRAS.430..767C}) and do not allow for HOD environmental variations. \cite{2015arXiv150607523Z} describe efficient procedures for exploring the HOD parameter space, which will be useful for this daunting computational task.  GGL measurements from the SDSS main galaxy sample already yield tight constraints on the galaxy-halo connection for fixed cosmological parameters (\citealt{2015MNRAS.454.1161Z}), and application of the methods described here could yield competitive new constraints on the amplitude of low redshift matter clustering.  On the several year timescale, measurements from the Dark Energy Survey and the Subaru Hyper-Suprime Camera could easily tighten these constraints to the one-percent level.  The weak lensing surveys of LSST, Euclid, and WFIRST seek a further order-of-magnitude improvement in measurement precision, presenting a stiff challenge for theoretical modeling and a superb opportunity to test our understanding of dark energy and gravity on cosmological scales. 
\section{Acknowledgements} 
We thank Andrew Hearin and Doug Watson for providing their scrambled catalogs and for useful comments and discussion.  We benefited more broadly from discussions with many participants at the 2014 CCAPP workshop on Assembly Bias and the Galaxy-Halo Relation.  We thank Ben Wibking for insightful conversations and valuable technical advice and Ashley Ross for discussions of the observational aspects of HOD analysis.This work was supported by NSF grant AST1516997.

\bibliographystyle{mnras}
\bibliography{ref}

\begin{thebibliography}{}
\makeatletter
\relax
\def\mn@urlcharsother{\let\do\@makeother \do\$\do\&\do\#\do\^\do\_\do\%\do\~}
\def\mn@doi{\begingroup\mn@urlcharsother \@ifnextchar [ {\mn@doi@}
  {\mn@doi@[]}}
\def\mn@doi@[#1]#2{\def\@tempa{#1}\ifx\@tempa\@empty \href
  {http://dx.doi.org/#2} {doi:#2}\else \href {http://dx.doi.org/#2} {#1}\fi
  \endgroup}
\def\mn@eprint#1#2{\mn@eprint@#1:#2::\@nil}
\def\mn@eprint@arXiv#1{\href {http://arxiv.org/abs/#1} {{\tt arXiv:#1}}}
\def\mn@eprint@dblp#1{\href {http://dblp.uni-trier.de/rec/bibtex/#1.xml}
  {dblp:#1}}
\def\mn@eprint@#1:#2:#3:#4\@nil{\def\@tempa {#1}\def\@tempb {#2}\def\@tempc
  {#3}\ifx \@tempc \@empty \let \@tempc \@tempb \let \@tempb \@tempa \fi \ifx
  \@tempb \@empty \def\@tempb {arXiv}\fi \@ifundefined
  {mn@eprint@\@tempb}{\@tempb:\@tempc}{\expandafter \expandafter \csname
  mn@eprint@\@tempb\endcsname \expandafter{\@tempc}}}

\bibitem[\protect\citeauthoryear{{Angulo}, {Baugh}  \& {Lacey}}{{Angulo}
  et~al.}{2008}]{2008MNRAS.387..921A}
{Angulo} R.~E.,  {Baugh} C.~M.,   {Lacey} C.~G.,  2008, \mn@doi [\mnras]
  {10.1111/j.1365-2966.2008.13304.x}, \href
  {http://adsabs.harvard.edu/abs/2008MNRAS.387..921A} {387, 921}

\bibitem[\protect\citeauthoryear{{Aubourg} et~al.,}{{Aubourg}
  et~al.}{2014}]{2014arXiv1411.1074A}
{Aubourg} {\'E}.,  et~al., 2014, ArXiv:1411.1074A, \href
  {http://adsabs.harvard.edu/abs/2014arXiv1411.1074A} {}

\bibitem[\protect\citeauthoryear{{Avila-Reese}, {Col{\'{\i}}n},
  {Gottl{\"o}ber}, {Firmani}  \& {Maulbetsch}}{{Avila-Reese}
  et~al.}{2005}]{2005ApJ...634...51A}
{Avila-Reese} V.,  {Col{\'{\i}}n} P.,  {Gottl{\"o}ber} S.,  {Firmani} C.,
  {Maulbetsch} C.,  2005, \mn@doi [\apj] {10.1086/491726}, \href
  {http://adsabs.harvard.edu/abs/2005ApJ...634...51A} {634, 51}

\bibitem[\protect\citeauthoryear{{Baldauf}, {Smith}, {Seljak}  \&
  {Mandelbaum}}{{Baldauf} et~al.}{2010}]{2010PhRvD..81f3531B}
{Baldauf} T.,  {Smith} R.~E.,  {Seljak} U.,   {Mandelbaum} R.,  2010, \mn@doi
  [\prd] {10.1103/PhysRevD.81.063531}, \href
  {http://adsabs.harvard.edu/abs/2010PhRvD..81f3531B} {81, 063531}

\bibitem[\protect\citeauthoryear{{Bartelmann} \& {Schneider}}{{Bartelmann} \&
  {Schneider}}{2001}]{2001PhR...340..291B}
{Bartelmann} M.,  {Schneider} P.,  2001, \mn@doi [Physics Reports]
  {10.1016/S0370-1573(00)00082-X}, \href
  {http://adsabs.harvard.edu/abs/2001PhR...340..291B} {340, 291}

\bibitem[\protect\citeauthoryear{{Behroozi}, {Wechsler}, {Wu}, {Busha},
  {Klypin}  \& {Primack}}{{Behroozi} et~al.}{2013}]{2013ApJ...763...18B}
{Behroozi} P.~S.,  {Wechsler} R.~H.,  {Wu} H.-Y.,  {Busha} M.~T.,  {Klypin}
  A.~A.,   {Primack} J.~R.,  2013, \mn@doi [\apj] {10.1088/0004-637X/763/1/18},
  \href {http://adsabs.harvard.edu/abs/2013ApJ...763...18B} {763, 18}

\bibitem[\protect\citeauthoryear{{Benson}, {Cole}, {Frenk}, {Baugh}  \&
  {Lacey}}{{Benson} et~al.}{2000}]{2000MNRAS.311..793B}
{Benson} A.~J.,  {Cole} S.,  {Frenk} C.~S.,  {Baugh} C.~M.,   {Lacey} C.~G.,
  2000, \mn@doi [\mnras] {10.1046/j.1365-8711.2000.03101.x}, \href
  {http://adsabs.harvard.edu/abs/2000MNRAS.311..793B} {311, 793}

\bibitem[\protect\citeauthoryear{{Berlind} \& {Weinberg}}{{Berlind} \&
  {Weinberg}}{2002}]{2002ApJ...575..587B}
{Berlind} A.~A.,  {Weinberg} D.~H.,  2002, \mn@doi [\apj] {10.1086/341469},
  \href {http://adsabs.harvard.edu/abs/2002ApJ...575..587B} {575, 587}

\bibitem[\protect\citeauthoryear{{Berlind} et~al.,}{{Berlind}
  et~al.}{2003}]{2003ApJ...593....1B}
{Berlind} A.~A.,  et~al., 2003, \mn@doi [\apj] {10.1086/376517}, \href
  {http://adsabs.harvard.edu/abs/2003ApJ...593....1B} {593, 1}

\bibitem[\protect\citeauthoryear{{Bett}, {Eke}, {Frenk}, {Jenkins}, {Helly}  \&
  {Navarro}}{{Bett} et~al.}{2007}]{2007MNRAS.376..215B}
{Bett} P.,  {Eke} V.,  {Frenk} C.~S.,  {Jenkins} A.,  {Helly} J.,   {Navarro}
  J.,  2007, \mn@doi [\mnras] {10.1111/j.1365-2966.2007.11432.x}, \href
  {http://adsabs.harvard.edu/abs/2007MNRAS.376..215B} {376, 215}

\bibitem[\protect\citeauthoryear{{Bond}, {Cole}, {Efstathiou}  \&
  {Kaiser}}{{Bond} et~al.}{1991}]{1991ApJ...379..440B}
{Bond} J.~R.,  {Cole} S.,  {Efstathiou} G.,   {Kaiser} N.,  1991, \mn@doi
  [\apj] {10.1086/170520}, \href
  {http://adsabs.harvard.edu/abs/1991ApJ...379..440B} {379, 440}

\bibitem[\protect\citeauthoryear{{Cacciato}, {van den Bosch}, {More}, {Li},
  {Mo}  \& {Yang}}{{Cacciato} et~al.}{2009}]{2009MNRAS.394..929C}
{Cacciato} M.,  {van den Bosch} F.~C.,  {More} S.,  {Li} R.,  {Mo} H.~J.,
  {Yang} X.,  2009, \mn@doi [\mnras] {10.1111/j.1365-2966.2008.14362.x}, \href
  {http://adsabs.harvard.edu/abs/2009MNRAS.394..929C} {394, 929}

\bibitem[\protect\citeauthoryear{{Cacciato}, {Lahav}, {van den Bosch},
  {Hoekstra}  \& {Dekel}}{{Cacciato} et~al.}{2012}]{2012MNRAS.426..566C}
{Cacciato} M.,  {Lahav} O.,  {van den Bosch} F.~C.,  {Hoekstra} H.,   {Dekel}
  A.,  2012, \mn@doi [\mnras] {10.1111/j.1365-2966.2012.21762.x}, \href
  {http://adsabs.harvard.edu/abs/2012MNRAS.426..566C} {426, 566}

\bibitem[\protect\citeauthoryear{{Cacciato}, {van den Bosch}, {More}, {Mo}  \&
  {Yang}}{{Cacciato} et~al.}{2013}]{2013MNRAS.430..767C}
{Cacciato} M.,  {van den Bosch} F.~C.,  {More} S.,  {Mo} H.,   {Yang} X.,
  2013, \mn@doi [\mnras] {10.1093/mnras/sts525}, \href
  {http://adsabs.harvard.edu/abs/2013MNRAS.430..767C} {430, 767}

\bibitem[\protect\citeauthoryear{{Chaves-Montero}, {Angulo}, {Schaye},
  {Schaller}, {Crain}  \& {Furlong}}{{Chaves-Montero}
  et~al.}{2015}]{2015arXiv150701948C}
{Chaves-Montero} J.,  {Angulo} R.~E.,  {Schaye} J.,  {Schaller} M.,  {Crain}
  R.~A.,   {Furlong} M.,  2015, ArXiv:150701948C, \href
  {http://adsabs.harvard.edu/abs/2015arXiv150701948C} {}

\bibitem[\protect\citeauthoryear{{Conroy} \& {Wechsler}}{{Conroy} \&
  {Wechsler}}{2009}]{2009ApJ...696..620C}
{Conroy} C.,  {Wechsler} R.~H.,  2009, \mn@doi [\apj]
  {10.1088/0004-637X/696/1/620}, \href
  {http://adsabs.harvard.edu/abs/2009ApJ...696..620C} {696, 620}

\bibitem[\protect\citeauthoryear{{Conroy}, {Wechsler}  \& {Kravtsov}}{{Conroy}
  et~al.}{2006}]{2006ApJ...647..201C}
{Conroy} C.,  {Wechsler} R.~H.,   {Kravtsov} A.~V.,  2006, \mn@doi [\apj]
  {10.1086/503602}, \href {http://adsabs.harvard.edu/abs/2006ApJ...647..201C}
  {647, 201}

\bibitem[\protect\citeauthoryear{{Coupon} et~al.,}{{Coupon}
  et~al.}{2012}]{2012A&A...542A...5C}
{Coupon} J.,  et~al., 2012, \mn@doi [\aap] {10.1051/0004-6361/201117625}, \href
  {http://adsabs.harvard.edu/abs/2012A%26A...542A...5C} {542, A5}

\bibitem[\protect\citeauthoryear{{Coupon} et~al.,}{{Coupon}
  et~al.}{2015}]{2015MNRAS.449.1352C}
{Coupon} J.,  et~al., 2015, \mn@doi [\mnras] {10.1093/mnras/stv276}, \href
  {http://adsabs.harvard.edu/abs/2015MNRAS.449.1352C} {449, 1352}

\bibitem[\protect\citeauthoryear{{Croton}, {Gao}  \& {White}}{{Croton}
  et~al.}{2007}]{2007MNRAS.374.1303C}
{Croton} D.~J.,  {Gao} L.,   {White} S.~D.~M.,  2007, \mn@doi [\mnras]
  {10.1111/j.1365-2966.2006.11230.x}, \href
  {http://adsabs.harvard.edu/abs/2007MNRAS.374.1303C} {374, 1303}

\bibitem[\protect\citeauthoryear{{Dalal}, {White}, {Bond}  \&
  {Shirokov}}{{Dalal} et~al.}{2008}]{2008ApJ...687...12D}
{Dalal} N.,  {White} M.,  {Bond} J.~R.,   {Shirokov} A.,  2008, \mn@doi [\apj]
  {10.1086/591512}, \href {http://adsabs.harvard.edu/abs/2008ApJ...687...12D}
  {687, 12}

\bibitem[\protect\citeauthoryear{{Dawson} et~al.,}{{Dawson}
  et~al.}{2013}]{2013AJ....145...10D}
{Dawson} K.~S.,  et~al., 2013, \mn@doi [\aj] {10.1088/0004-6256/145/1/10},
  \href {http://adsabs.harvard.edu/abs/2013AJ....145...10D} {145, 10}

\bibitem[\protect\citeauthoryear{{Eisenstein} et~al.,}{{Eisenstein}
  et~al.}{2001}]{2001AJ....122.2267E}
{Eisenstein} D.~J.,  et~al., 2001, \mn@doi [\aj] {10.1086/323717}, \href
  {http://adsabs.harvard.edu/abs/2001AJ....122.2267E} {122, 2267}

\bibitem[\protect\citeauthoryear{{Fakhouri} \& {Ma}}{{Fakhouri} \&
  {Ma}}{2009}]{2009MNRAS.394.1825F}
{Fakhouri} O.,  {Ma} C.-P.,  2009, \mn@doi [\mnras]
  {10.1111/j.1365-2966.2009.14480.x}, \href
  {http://adsabs.harvard.edu/abs/2009MNRAS.394.1825F} {394, 1825}

\bibitem[\protect\citeauthoryear{{Faltenbacher} \& {White}}{{Faltenbacher} \&
  {White}}{2010}]{2010ApJ...708..469F}
{Faltenbacher} A.,  {White} S.~D.~M.,  2010, \mn@doi [\apj]
  {10.1088/0004-637X/708/1/469}, \href
  {http://adsabs.harvard.edu/abs/2010ApJ...708..469F} {708, 469}

\bibitem[\protect\citeauthoryear{{Frieman}, {Turner}  \& {Huterer}}{{Frieman}
  et~al.}{2008}]{2008ARA&A..46..385F}
{Frieman} J.~A.,  {Turner} M.~S.,   {Huterer} D.,  2008, \mn@doi [ARA&A]
  {10.1146/annurev.astro.46.060407.145243}, \href
  {http://adsabs.harvard.edu/abs/2008ARA%26A..46..385F} {46, 385}

\bibitem[\protect\citeauthoryear{{Gao}, {Springel}  \& {White}}{{Gao}
  et~al.}{2005}]{2005MNRAS.363L..66G}
{Gao} L.,  {Springel} V.,   {White} S.~D.~M.,  2005, \mn@doi [\mnras]
  {10.1111/j.1745-3933.2005.00084.x}, \href
  {http://adsabs.harvard.edu/abs/2005MNRAS.363L..66G} {363, L66}

\bibitem[\protect\citeauthoryear{{Gottl{\"o}ber}, {Klypin}  \&
  {Kravtsov}}{{Gottl{\"o}ber} et~al.}{2001}]{2001PABei..19S..58G}
{Gottl{\"o}ber} S.,  {Klypin} A.,   {Kravtsov} A.~V.,  2001, Progress in
  Astronomy, \href {http://adsabs.harvard.edu/abs/2001PABei..19S..58G} {19, 58}

\bibitem[\protect\citeauthoryear{{Gottloeber} \& {Klypin}}{{Gottloeber} \&
  {Klypin}}{2008}]{2008arXiv0803.4343G}
{Gottloeber} S.,  {Klypin} A.,  2008, ArXiv:0803.4343G, \href
  {http://adsabs.harvard.edu/abs/2008arXiv0803.4343G} {}

\bibitem[\protect\citeauthoryear{{Guo}, {White}, {Li}  \&
  {Boylan-Kolchin}}{{Guo} et~al.}{2010}]{2010MNRAS.404.1111G}
{Guo} Q.,  {White} S.,  {Li} C.,   {Boylan-Kolchin} M.,  2010, \mn@doi [\mnras]
  {10.1111/j.1365-2966.2010.16341.x}, \href
  {http://adsabs.harvard.edu/abs/2010MNRAS.404.1111G} {404, 1111}

\bibitem[\protect\citeauthoryear{{Guzik} \& {Seljak}}{{Guzik} \&
  {Seljak}}{2001}]{2001MNRAS.321..439G}
{Guzik} J.,  {Seljak} U.,  2001, \mn@doi [\mnras]
  {10.1046/j.1365-8711.2001.04081.x}, \href
  {http://adsabs.harvard.edu/abs/2001MNRAS.321..439G} {321, 439}

\bibitem[\protect\citeauthoryear{{Harker}, {Cole}, {Helly}, {Frenk}  \&
  {Jenkins}}{{Harker} et~al.}{2006}]{2006MNRAS.367.1039H}
{Harker} G.,  {Cole} S.,  {Helly} J.,  {Frenk} C.,   {Jenkins} A.,  2006,
  \mn@doi [\mnras] {10.1111/j.1365-2966.2006.10022.x}, \href
  {http://adsabs.harvard.edu/abs/2006MNRAS.367.1039H} {367, 1039}

\bibitem[\protect\citeauthoryear{{Hearin} \& {Watson}}{{Hearin} \&
  {Watson}}{2013}]{2013MNRAS.435.1313H}
{Hearin} A.~P.,  {Watson} D.~F.,  2013, \mn@doi [\mnras]
  {10.1093/mnras/stt1374}, \href
  {http://adsabs.harvard.edu/abs/2013MNRAS.435.1313H} {435, 1313}

\bibitem[\protect\citeauthoryear{{Hearin}, {Watson}, {Becker}, {Reyes},
  {Berlind}  \& {Zentner}}{{Hearin} et~al.}{2014}]{2014MNRAS.444..729H}
{Hearin} A.~P.,  {Watson} D.~F.,  {Becker} M.~R.,  {Reyes} R.,  {Berlind}
  A.~A.,   {Zentner} A.~R.,  2014, \mn@doi [\mnras] {10.1093/mnras/stu1443},
  \href {http://adsabs.harvard.edu/abs/2014MNRAS.444..729H} {444, 729}

\bibitem[\protect\citeauthoryear{{Hearin}, {Zentner}, {van den Bosch},
  {Campbell}  \& {Tollerud}}{{Hearin} et~al.}{2015}]{2015arXiv151203050H}
{Hearin} A.~P.,  {Zentner} A.~R.,  {van den Bosch} F.~C.,  {Campbell} D.,
  {Tollerud} E.,  2015, arXiv: 151203050, \href
  {http://adsabs.harvard.edu/abs/2015arXiv151203050H} {}

\bibitem[\protect\citeauthoryear{{Heymans} et~al.,}{{Heymans}
  et~al.}{2012}]{2012MNRAS.427..146H}
{Heymans} C.,  et~al., 2012, \mn@doi [\mnras]
  {10.1111/j.1365-2966.2012.21952.x}, \href
  {http://adsabs.harvard.edu/abs/2012MNRAS.427..146H} {427, 146}

\bibitem[\protect\citeauthoryear{{Jing}, {Mo}  \& {B{\"o}rner}}{{Jing}
  et~al.}{1998}]{1998ApJ...494....1J}
{Jing} Y.~P.,  {Mo} H.~J.,   {B{\"o}rner} G.,  1998, \mn@doi [\apj]
  {10.1086/305209}, \href {http://adsabs.harvard.edu/abs/1998ApJ...494....1J}
  {494, 1}

\bibitem[\protect\citeauthoryear{{Jing}, {Suto}  \& {Mo}}{{Jing}
  et~al.}{2007}]{2007ApJ...657..664J}
{Jing} Y.~P.,  {Suto} Y.,   {Mo} H.~J.,  2007, \mn@doi [\apj] {10.1086/511130},
  \href {http://adsabs.harvard.edu/abs/2007ApJ...657..664J} {657, 664}

\bibitem[\protect\citeauthoryear{{Klypin}, {Trujillo-Gomez}  \&
  {Primack}}{{Klypin} et~al.}{2011}]{2011ApJ...740..102K}
{Klypin} A.~A.,  {Trujillo-Gomez} S.,   {Primack} J.,  2011, \mn@doi [\apj]
  {10.1088/0004-637X/740/2/102}, \href
  {http://adsabs.harvard.edu/abs/2011ApJ...740..102K} {740, 102}

\bibitem[\protect\citeauthoryear{{Kravtsov}}{{Kravtsov}}{2013}]{2013ApJ...764L..31K}
{Kravtsov} A.~V.,  2013, \mn@doi [\apjl] {10.1088/2041-8205/764/2/L31}, \href
  {http://adsabs.harvard.edu/abs/2013ApJ...764L..31K} {764, L31}

\bibitem[\protect\citeauthoryear{{Kravtsov}, {Klypin}  \&
  {Khokhlov}}{{Kravtsov} et~al.}{1997}]{1997ApJS..111...73K}
{Kravtsov} A.~V.,  {Klypin} A.~A.,   {Khokhlov} A.~M.,  1997, \mn@doi [\apjs]
  {10.1086/313015}, \href {http://adsabs.harvard.edu/abs/1997ApJS..111...73K}
  {111, 73}

\bibitem[\protect\citeauthoryear{{Kravtsov}, {Berlind}, {Wechsler}, {Klypin},
  {Gottl{\"o}ber}, {Allgood}  \& {Primack}}{{Kravtsov}
  et~al.}{2004}]{2004ApJ...609...35K}
{Kravtsov} A.~V.,  {Berlind} A.~A.,  {Wechsler} R.~H.,  {Klypin} A.~A.,
  {Gottl{\"o}ber} S.,  {Allgood} B.,   {Primack} J.~R.,  2004, \mn@doi [\apj]
  {10.1086/420959}, \href {http://adsabs.harvard.edu/abs/2004ApJ...609...35K}
  {609, 35}

\bibitem[\protect\citeauthoryear{{Leauthaud}, {Tinker}, {Behroozi}, {Busha}  \&
  {Wechsler}}{{Leauthaud} et~al.}{2011}]{2011ApJ...738...45L}
{Leauthaud} A.,  {Tinker} J.,  {Behroozi} P.~S.,  {Busha} M.~T.,   {Wechsler}
  R.~H.,  2011, \mn@doi [\apj] {10.1088/0004-637X/738/1/45}, \href
  {http://adsabs.harvard.edu/abs/2011ApJ...738...45L} {738, 45}

\bibitem[\protect\citeauthoryear{{Ma} \& {Fry}}{{Ma} \&
  {Fry}}{2000}]{2000ApJ...531L..87M}
{Ma} C.-P.,  {Fry} J.~N.,  2000, \mn@doi [\apjl] {10.1086/312534}, \href
  {http://adsabs.harvard.edu/abs/2000ApJ...531L..87M} {531, L87}

\bibitem[\protect\citeauthoryear{{Mandelbaum}, {Slosar}, {Baldauf}, {Seljak},
  {Hirata}, {Nakajima}, {Reyes}  \& {Smith}}{{Mandelbaum}
  et~al.}{2013}]{2013MNRAS.432.1544M}
{Mandelbaum} R.,  {Slosar} A.,  {Baldauf} T.,  {Seljak} U.,  {Hirata} C.~M.,
  {Nakajima} R.,  {Reyes} R.,   {Smith} R.~E.,  2013, \mn@doi [\mnras]
  {10.1093/mnras/stt572}, \href
  {http://adsabs.harvard.edu/abs/2013MNRAS.432.1544M} {432, 1544}

\bibitem[\protect\citeauthoryear{{Maulbetsch}, {Avila-Reese}, {Col{\'{\i}}n},
  {Gottl{\"o}ber}, {Khalatyan}  \& {Steinmetz}}{{Maulbetsch}
  et~al.}{2007}]{2007ApJ...654...53M}
{Maulbetsch} C.,  {Avila-Reese} V.,  {Col{\'{\i}}n} P.,  {Gottl{\"o}ber} S.,
  {Khalatyan} A.,   {Steinmetz} M.,  2007, \mn@doi [\apj] {10.1086/509706},
  \href {http://adsabs.harvard.edu/abs/2007ApJ...654...53M} {654, 53}

\bibitem[\protect\citeauthoryear{{Mehta}}{{Mehta}}{2014}]{2014PhDT.......126M}
{Mehta} K.~T.,  2014, PhD thesis, The University of Arizona

\bibitem[\protect\citeauthoryear{{More}, {Miyatake}, {Mandelbaum}, {Takada},
  {Spergel}, {Brownstein}  \& {Schneider}}{{More}
  et~al.}{2015}]{2015ApJ...806....2M}
{More} S.,  {Miyatake} H.,  {Mandelbaum} R.,  {Takada} M.,  {Spergel} D.~N.,
  {Brownstein} J.~R.,   {Schneider} D.~P.,  2015, \mn@doi [\apj]
  {10.1088/0004-637X/806/1/2}, \href
  {http://adsabs.harvard.edu/abs/2015ApJ...806....2M} {806, 2}

\bibitem[\protect\citeauthoryear{{Mortonson}, {Weinberg}  \&
  {White}}{{Mortonson} et~al.}{2014}]{2014arXiv1401.0046M}
{Mortonson} M.~J.,  {Weinberg} D.~H.,   {White} M.,  2014, Partilce Data Group
  review on Dark Eenrgy available as arXiv:1401.0046, \href
  {http://adsabs.harvard.edu/abs/2014arXiv1401.0046M} {}

\bibitem[\protect\citeauthoryear{{Nagai} \& {Kravtsov}}{{Nagai} \&
  {Kravtsov}}{2005}]{2005ApJ...618..557N}
{Nagai} D.,  {Kravtsov} A.~V.,  2005, \mn@doi [\apj] {10.1086/426016}, \href
  {http://adsabs.harvard.edu/abs/2005ApJ...618..557N} {618, 557}

\bibitem[\protect\citeauthoryear{{Navarro}, {Frenk}  \& {White}}{{Navarro}
  et~al.}{1997}]{1997ApJ...490..493N}
{Navarro} J.~F.,  {Frenk} C.~S.,   {White} S.~D.~M.,  1997, \apj, \href
  {http://adsabs.harvard.edu/abs/1997ApJ...490..493N} {490, 493}

\bibitem[\protect\citeauthoryear{{Neistein}, {Weinmann}, {Li}  \&
  {Boylan-Kolchin}}{{Neistein} et~al.}{2011}]{2011MNRAS.414.1405N}
{Neistein} E.,  {Weinmann} S.~M.,  {Li} C.,   {Boylan-Kolchin} M.,  2011,
  \mn@doi [\mnras] {10.1111/j.1365-2966.2011.18473.x}, \href
  {http://adsabs.harvard.edu/abs/2011MNRAS.414.1405N} {414, 1405}

\bibitem[\protect\citeauthoryear{{Peacock} \& {Smith}}{{Peacock} \&
  {Smith}}{2000}]{2000MNRAS.318.1144P}
{Peacock} J.~A.,  {Smith} R.~E.,  2000, \mn@doi [\mnras]
  {10.1046/j.1365-8711.2000.03779.x}, \href
  {http://adsabs.harvard.edu/abs/2000MNRAS.318.1144P} {318, 1144}

\bibitem[\protect\citeauthoryear{{Planck Collaboration}}{{Planck
  Collaboration}}{2015}]{2015arXiv150201597P}
{Planck Collaboration} 2015, ArXiv:150201597P, \href
  {http://adsabs.harvard.edu/abs/2015arXiv150201597P} {}

\bibitem[\protect\citeauthoryear{{Reddick}, {Wechsler}, {Tinker}  \&
  {Behroozi}}{{Reddick} et~al.}{2013}]{2013ApJ...771...30R}
{Reddick} R.~M.,  {Wechsler} R.~H.,  {Tinker} J.~L.,   {Behroozi} P.~S.,  2013,
  \mn@doi [\apj] {10.1088/0004-637X/771/1/30}, \href
  {http://adsabs.harvard.edu/abs/2013ApJ...771...30R} {771, 30}

\bibitem[\protect\citeauthoryear{{Rodr{\'{\i}}guez-Puebla}, {Drory}  \&
  {Avila-Reese}}{{Rodr{\'{\i}}guez-Puebla} et~al.}{2012}]{2012ApJ...756....2R}
{Rodr{\'{\i}}guez-Puebla} A.,  {Drory} N.,   {Avila-Reese} V.,  2012, \mn@doi
  [\apj] {10.1088/0004-637X/756/1/2}, \href
  {http://adsabs.harvard.edu/abs/2012ApJ...756....2R} {756, 2}

\bibitem[\protect\citeauthoryear{{Rozo} et~al.,}{{Rozo}
  et~al.}{2015}]{2015arXiv150705460R}
{Rozo} E.,  et~al., 2015, ArXiv:150705460R, \href
  {http://adsabs.harvard.edu/abs/2015arXiv150705460R} {}

\bibitem[\protect\citeauthoryear{{Schaye} et~al.,}{{Schaye}
  et~al.}{2015}]{2015MNRAS.446..521S}
{Schaye} J.,  et~al., 2015, \mn@doi [\mnras] {10.1093/mnras/stu2058}, \href
  {http://adsabs.harvard.edu/abs/2015MNRAS.446..521S} {446, 521}

\bibitem[\protect\citeauthoryear{{Scoccimarro}, {Sheth}, {Hui}  \&
  {Jain}}{{Scoccimarro} et~al.}{2001}]{2001ApJ...546...20S}
{Scoccimarro} R.,  {Sheth} R.~K.,  {Hui} L.,   {Jain} B.,  2001, \mn@doi [\apj]
  {10.1086/318261}, \href {http://adsabs.harvard.edu/abs/2001ApJ...546...20S}
  {546, 20}

\bibitem[\protect\citeauthoryear{{Seljak}}{{Seljak}}{2000}]{2000MNRAS.318..203S}
{Seljak} U.,  2000, \mn@doi [\mnras] {10.1046/j.1365-8711.2000.03715.x}, \href
  {http://adsabs.harvard.edu/abs/2000MNRAS.318..203S} {318, 203}

\bibitem[\protect\citeauthoryear{{Sheldon} et~al.,}{{Sheldon}
  et~al.}{2004}]{2004AJ....127.2544S}
{Sheldon} E.~S.,  et~al., 2004, \mn@doi [\aj] {10.1086/383293}, \href
  {http://adsabs.harvard.edu/abs/2004AJ....127.2544S} {127, 2544}

\bibitem[\protect\citeauthoryear{{Sheth} \& {Tormen}}{{Sheth} \&
  {Tormen}}{2004}]{2004MNRAS.350.1385S}
{Sheth} R.~K.,  {Tormen} G.,  2004, \mn@doi [\mnras]
  {10.1111/j.1365-2966.2004.07733.x}, \href
  {http://adsabs.harvard.edu/abs/2004MNRAS.350.1385S} {350, 1385}

\bibitem[\protect\citeauthoryear{{Simha}, {Weinberg}, {Dav{\'e}}, {Gnedin},
  {Katz}  \& {Kere{\v s}}}{{Simha} et~al.}{2009}]{2009MNRAS.399..650S}
{Simha} V.,  {Weinberg} D.~H.,  {Dav{\'e}} R.,  {Gnedin} O.~Y.,  {Katz} N.,
  {Kere{\v s}} D.,  2009, \mn@doi [\mnras] {10.1111/j.1365-2966.2009.15341.x},
  \href {http://adsabs.harvard.edu/abs/2009MNRAS.399..650S} {399, 650}

\bibitem[\protect\citeauthoryear{{Simha}, {Weinberg}, {Dav{\'e}}, {Fardal},
  {Katz}  \& {Oppenheimer}}{{Simha} et~al.}{2012}]{2012MNRAS.423.3458S}
{Simha} V.,  {Weinberg} D.~H.,  {Dav{\'e}} R.,  {Fardal} M.,  {Katz} N.,
  {Oppenheimer} B.~D.,  2012, \mn@doi [\mnras]
  {10.1111/j.1365-2966.2012.21142.x}, \href
  {http://adsabs.harvard.edu/abs/2012MNRAS.423.3458S} {423, 3458}

\bibitem[\protect\citeauthoryear{{Strauss} et~al.,}{{Strauss}
  et~al.}{2002}]{2002AJ....124.1810S}
{Strauss} M.~A.,  et~al., 2002, \mn@doi [\aj] {10.1086/342343}, \href
  {http://adsabs.harvard.edu/abs/2002AJ....124.1810S} {124, 1810}

\bibitem[\protect\citeauthoryear{{Tasitsiomi}, {Kravtsov}, {Wechsler}  \&
  {Primack}}{{Tasitsiomi} et~al.}{2004}]{2004ApJ...614..533T}
{Tasitsiomi} A.,  {Kravtsov} A.~V.,  {Wechsler} R.~H.,   {Primack} J.~R.,
  2004, \mn@doi [\apj] {10.1086/423784}, \href
  {http://adsabs.harvard.edu/abs/2004ApJ...614..533T} {614, 533}

\bibitem[\protect\citeauthoryear{{Vale} \& {Ostriker}}{{Vale} \&
  {Ostriker}}{2004}]{2004MNRAS.353..189V}
{Vale} A.,  {Ostriker} J.~P.,  2004, \mn@doi [\mnras]
  {10.1111/j.1365-2966.2004.08059.x}, \href
  {http://adsabs.harvard.edu/abs/2004MNRAS.353..189V} {353, 189}

\bibitem[\protect\citeauthoryear{{Wang}, {Mo}  \& {Jing}}{{Wang}
  et~al.}{2007}]{2007MNRAS.375..633W}
{Wang} H.~Y.,  {Mo} H.~J.,   {Jing} Y.~P.,  2007, \mn@doi [\mnras]
  {10.1111/j.1365-2966.2006.11316.x}, \href
  {http://adsabs.harvard.edu/abs/2007MNRAS.375..633W} {375, 633}

\bibitem[\protect\citeauthoryear{{Watson}, {Berlind}  \& {Zentner}}{{Watson}
  et~al.}{2012}]{2012ApJ...754...90W}
{Watson} D.~F.,  {Berlind} A.~A.,   {Zentner} A.~R.,  2012, \mn@doi [\apj]
  {10.1088/0004-637X/754/2/90}, \href
  {http://adsabs.harvard.edu/abs/2012ApJ...754...90W} {754, 90}

\bibitem[\protect\citeauthoryear{{Wechsler}, {Bullock}, {Primack}, {Kravtsov}
  \& {Dekel}}{{Wechsler} et~al.}{2002}]{2002ApJ...568...52W}
{Wechsler} R.~H.,  {Bullock} J.~S.,  {Primack} J.~R.,  {Kravtsov} A.~V.,
  {Dekel} A.,  2002, \mn@doi [\apj] {10.1086/338765}, \href
  {http://adsabs.harvard.edu/abs/2002ApJ...568...52W} {568, 52}

\bibitem[\protect\citeauthoryear{{Wechsler}, {Zentner}, {Bullock}, {Kravtsov}
  \& {Allgood}}{{Wechsler} et~al.}{2006}]{2006ApJ...652...71W}
{Wechsler} R.~H.,  {Zentner} A.~R.,  {Bullock} J.~S.,  {Kravtsov} A.~V.,
  {Allgood} B.,  2006, \mn@doi [\apj] {10.1086/507120}, \href
  {http://adsabs.harvard.edu/abs/2006ApJ...652...71W} {652, 71}

\bibitem[\protect\citeauthoryear{{Weinberg}, {Mortonson}, {Eisenstein},
  {Hirata}, {Riess}  \& {Rozo}}{{Weinberg} et~al.}{2013}]{2013PhR...530...87W}
{Weinberg} D.~H.,  {Mortonson} M.~J.,  {Eisenstein} D.~J.,  {Hirata} C.,
  {Riess} A.~G.,   {Rozo} E.,  2013, \mn@doi [\physrep]
  {10.1016/j.physrep.2013.05.001}, \href
  {http://adsabs.harvard.edu/abs/2013PhR...530...87W} {530, 87}

\bibitem[\protect\citeauthoryear{{Wetzel}, {Cohn}, {White}, {Holz}  \&
  {Warren}}{{Wetzel} et~al.}{2007}]{2007ApJ...656..139W}
{Wetzel} A.~R.,  {Cohn} J.~D.,  {White} M.,  {Holz} D.~E.,   {Warren} M.~S.,
  2007, \mn@doi [\apj] {10.1086/510444}, \href
  {http://adsabs.harvard.edu/abs/2007ApJ...656..139W} {656, 139}

\bibitem[\protect\citeauthoryear{{Yoo} \& {Seljak}}{{Yoo} \&
  {Seljak}}{2012}]{2012PhRvD..86h3504Y}
{Yoo} J.,  {Seljak} U.,  2012, \mn@doi [\prd] {10.1103/PhysRevD.86.083504},
  \href {http://adsabs.harvard.edu/abs/2012PhRvD..86h3504Y} {86, 083504}

\bibitem[\protect\citeauthoryear{{Yoo}, {Tinker}, {Weinberg}, {Zheng}, {Katz}
  \& {Dav{\'e}}}{{Yoo} et~al.}{2006}]{2006ApJ...652...26Y}
{Yoo} J.,  {Tinker} J.~L.,  {Weinberg} D.~H.,  {Zheng} Z.,  {Katz} N.,
  {Dav{\'e}} R.,  2006, \mn@doi [\apj] {10.1086/507591}, \href
  {http://adsabs.harvard.edu/abs/2006ApJ...652...26Y} {652, 26}

\bibitem[\protect\citeauthoryear{{Zehavi} et~al.,}{{Zehavi}
  et~al.}{2005}]{2005ApJ...630....1Z}
{Zehavi} I.,  et~al., 2005, \mn@doi [\apj] {10.1086/431891}, \href
  {http://adsabs.harvard.edu/abs/2005ApJ...630....1Z} {630, 1}

\bibitem[\protect\citeauthoryear{{Zehavi} et~al.,}{{Zehavi}
  et~al.}{2011}]{2011ApJ...736...59Z}
{Zehavi} I.,  et~al., 2011, \mn@doi [\apj] {10.1088/0004-637X/736/1/59}, \href
  {http://adsabs.harvard.edu/abs/2011ApJ...736...59Z} {736, 59}

\bibitem[\protect\citeauthoryear{{Zentner}, {Berlind}, {Bullock}, {Kravtsov}
  \& {Wechsler}}{{Zentner} et~al.}{2005}]{2005ApJ...624..505Z}
{Zentner} A.~R.,  {Berlind} A.~A.,  {Bullock} J.~S.,  {Kravtsov} A.~V.,
  {Wechsler} R.~H.,  2005, \mn@doi [\apj] {10.1086/428898}, \href
  {http://adsabs.harvard.edu/abs/2005ApJ...624..505Z} {624, 505}

\bibitem[\protect\citeauthoryear{{Zentner}, {Hearin}  \& {van den
  Bosch}}{{Zentner} et~al.}{2014}]{2014MNRAS.443.3044Z}
{Zentner} A.~R.,  {Hearin} A.~P.,   {van den Bosch} F.~C.,  2014, \mn@doi
  [\mnras] {10.1093/mnras/stu1383}, \href
  {http://adsabs.harvard.edu/abs/2014MNRAS.443.3044Z} {443, 3044}

\bibitem[\protect\citeauthoryear{{Zheng} \& {Guo}}{{Zheng} \&
  {Guo}}{2015}]{2015arXiv150607523Z}
{Zheng} Z.,  {Guo} H.,  2015, ArXiv:150607523Z, \href
  {http://adsabs.harvard.edu/abs/2015arXiv150607523Z} {}

\bibitem[\protect\citeauthoryear{{Zheng} \& {Weinberg}}{{Zheng} \&
  {Weinberg}}{2007}]{2007ApJ...659....1Z}
{Zheng} Z.,  {Weinberg} D.~H.,  2007, \mn@doi [\apj] {10.1086/512151}, \href
  {http://adsabs.harvard.edu/abs/2007ApJ...659....1Z} {659, 1}

\bibitem[\protect\citeauthoryear{{Zheng} et~al.,}{{Zheng}
  et~al.}{2005}]{2005ApJ...633..791Z}
{Zheng} Z.,  et~al., 2005, \mn@doi [\apj] {10.1086/466510}, \href
  {http://adsabs.harvard.edu/abs/2005ApJ...633..791Z} {633, 791}

\bibitem[\protect\citeauthoryear{{Zu} \& {Mandelbaum}}{{Zu} \&
  {Mandelbaum}}{2015}]{2015MNRAS.454.1161Z}
{Zu} Y.,  {Mandelbaum} R.,  2015, \mn@doi [\mnras] {10.1093/mnras/stv2062},
  \href {http://adsabs.harvard.edu/abs/2015MNRAS.454.1161Z} {454, 1161}

\bibitem[\protect\citeauthoryear{{van Daalen}, {Schaye}, {McCarthy}, {Booth}
  \& {Dalla Vecchia}}{{van Daalen} et~al.}{2014}]{2014MNRAS.440.2997V}
{van Daalen} M.~P.,  {Schaye} J.,  {McCarthy} I.~G.,  {Booth} C.~M.,   {Dalla
  Vecchia} C.,  2014, \mn@doi [\mnras] {10.1093/mnras/stu482}, \href
  {http://adsabs.harvard.edu/abs/2014MNRAS.440.2997V} {440, 2997}

\bibitem[\protect\citeauthoryear{{van den Bosch}, {More}, {Cacciato}, {Mo}  \&
  {Yang}}{{van den Bosch} et~al.}{2013}]{2013MNRAS.430..725V}
{van den Bosch} F.~C.,  {More} S.,  {Cacciato} M.,  {Mo} H.,   {Yang} X.,
  2013, \mn@doi [\mnras] {10.1093/mnras/sts006}, \href
  {http://adsabs.harvard.edu/abs/2013MNRAS.430..725V} {430, 725}

\makeatother
\end{thebibliography}

\end{document}